\documentclass[floatfix,aps,rmp,reprint,amsmath,amssymb,graphicx,longbibliography,superscriptaddress]{revtex4-1}

\usepackage{graphicx}
\usepackage[svgnames]{xcolor}
\usepackage{bm}
\usepackage{slashed}
\usepackage[pdfencoding=auto,psdextra]{hyperref}
\hypersetup{colorlinks,%
citecolor=violet,%
filecolor=black,%
linkcolor=black,%
urlcolor=blue}

\usepackage[normalem]{ulem}

\allowdisplaybreaks

\begin{document}

\title{Colloquium: The Cosmic Dipole Anomaly}

\author{Nathan Secrest}
\affiliation{U.S.\ Naval Observatory,\unpenalty~3450 Massachusetts Ave NW,\unpenalty~Washington,\unpenalty~DC 20392-5420,\unpenalty~USA}

\author{Sebastian von Hausegger}
\affiliation{Rudolf Peierls Centre for Theoretical Physics,\unpenalty~University of Oxford,\unpenalty~Parks Road,\unpenalty~Oxford OX1 3PU,\unpenalty~UK}

\author{Mohamed Rameez}
\affiliation{Department of High Energy Physics,\unpenalty~Tata Institute of Fundamental Research,\unpenalty~Homi Bhabha Road,\unpenalty~Mumbai 400005,\unpenalty~India}

\author{Roya Mohayaee}
\affiliation{Institut d'Astrophysique de Paris (IAP),\unpenalty~CNRS/Sorbonne Universit\'e,\unpenalty~98 bis Bld Arago,\unpenalty~Paris 75014,\unpenalty~France}

\author{Subir Sarkar}
\email{Corresponding author: subir.sarkar@physics.ox.ac.uk}
\affiliation{Rudolf Peierls Centre for Theoretical Physics,\unpenalty~University of Oxford,\unpenalty~Parks Road,\unpenalty~Oxford OX1 3PU,\unpenalty~UK}

\date{\today}  
\begin{abstract}
The Cosmological Principle, which states that the Universe is homogeneous and isotropic (when averaged on large scales), is the foundational assumption of Friedmann-Lema\^{i}tre-Robertson-Walker (FLRW) cosmologies such as the current standard Lambda-Cold-Dark-Matter ($\Lambda$CDM) model. This simplification yields an exact solution to the Einstein field equations that relates space and time through a single time-dependent scale factor, which defines cosmological observables such as the Hubble parameter and the cosmological redshift. The validity of the Cosmological Principle, which underpins modern cosmology, can now be rigorously tested with the advent of large, nearly all-sky catalogs of radio galaxies and quasars. Surprisingly, the dipole anisotropy in the large-scale distribution of matter is found to be inconsistent with the expectation from kinematic aberration and Doppler-boosting effects in a perturbed FLRW universe, which is the standard interpretation of the observed dipole in the cosmic microwave background (CMB). Although the matter dipole agrees in direction with that of the CMB dipole, it is anomalously larger, demonstrating that either the rest frames in which matter and radiation appear isotropic are not the same or that there is an unexpected intrinsic anisotropy in at least one of them. This discrepancy now exceeds $5\sigma$ in significance. We review these recent findings, as well as the potential biases, systematic issues, and alternate interpretations that have been suggested to help alleviate the tension. We conclude that the cosmic dipole anomaly poses a serious challenge to FLRW cosmology, and the standard $\Lambda$CDM model in particular, as an adequate description of our Universe.  
\end{abstract}
\maketitle
\tableofcontents{}

\section{The Cosmological Principle}
\label{sec:TheCosmologicalPrinciple} 

Between 1907 and 1915 Albert Einstein formulated the theory of General Relativity (GR), an elegant dynamical description of gravity as arising from the warping of space-time by mass-energy. The first discussions of how this may apply to the universe were presented soon afterwards \cite{Einstein:1917ce,deSitter:1917zz}. Alexander Friedmann provided the first solution of Einstein's equations applied to the universe as a whole --- which he predicted should be evolving \citep{Friedman:1922kd,Friedmann:1924bb}. However his work remained unknown, as also that of George Lema\^{i}tre who  first interpreted the redshifts that had been measured in the spectra of distant nebulae \citep{1917PAPhS..56..403S} in terms of the expansion of space itself, rather than the Doppler effect due to their relative motion \citep{Lemaitre:1927zz,Lemaitre:1933gd,Lemaitre:1931zzb}. Although just such a relationship between redshift (velocity) and distance was suggested subsequently by Edwin Hubble \cite{Hubble:1929ig}, it was not recognised as confirming the previous prediction. In fact Hubble  attempted to interpret the data in terms of the ``de Sitter effect" --- the dilation of time intervals for signals received from distant sources in an apparently static universe that de Sitter had presented. Subsequently when Hubble realised his error he wrote to de Sitter: “The interpretation, we feel, should be left to you and the very few others who are competent to discuss the matter with authority” \cite[see:][]{1993edhu.book.....S}. It is de Sitter's universe, which has in fact an accelerating rate of expansion, that is closest to today's standard $\Lambda$CDM cosmological model.

The mathematical model of the expanding universe constructed by all the above authors implicitly assumed what came to be called the Cosmological Principle (CP): ``The Universe must appear the same to all observers'' \cite{1932Natur.130....9M}. Karl Popper was unimpressed: “Because I dislike making of our lack of knowledge a principle of knowing something" \cite[see:][]{10.1162/POSC_a_00102}. Nevertheless this assumption is necessary, given that essentially all our information about the universe comes from within our past light cone; we cannot move over cosmological distances and check if the universe looks the same from over there \cite{1975QJRAS..16..245E}. Given the CP, the metric of space-time has the maximally symmetric Robertson-Walker form \cite{Robertson:1935zz,Walker:1937qxv}. This is the \emph{foundational} assumption that underpins today's standard $\Lambda$CDM cosmological model. 

Subsequently the CP was extended further to the `Perfect Cosmological Principle' which states that we have no special location in either space or time; this was the basis for the `Steady State theory' of the Universe \cite{1948MNRAS.108..252B}. The discovery of the cosmic microwave background (CMB) by \citet{Penzias:1965wn} established that the Universe was in fact different in the past and ruled out the Perfect Cosmological Principle. Its spatial version lived on however, in a modified form which acknowledges that the Universe is observed to be inhomogeneous, by asserting that it is the realisation of a spatially stationary stochastic process \cite{1962IAUS...15..294N,Peebles:1980yev}. In the standard $\Lambda$CDM cosmological model, the initial fluctuations which grew via gravitational instability into the observed large-scale structure are assumed to be a Gaussian random field, generated during a period of primordial inflation in the early Universe. The seed  fluctuations are tiny, of ${\cal O}(10^{-5})$, as observed via their imprint on the CMB. Hence although strictly speaking the Universe is not exactly isotropic or homogeneous on \emph{any} scale, observation of large-scale inhomogeneities or anisotropies, say a hundred times bigger would be quite adequate to falsify the CP. 

As \citet{Weinberg:1972kfs} stated in his influential textbook: 
\begin{quote}
The real reason, though, for our adherence here to the Cosmological Principle is not that it is surely correct, but rather, that it allows us to make use of the extremely limited data provided to cosmology by observational astronomy \ldots If the data will not fit into this framework, we shall be able to conclude that either the Cosmological Principle or the Principle of Equivalence is wrong. Nothing could be more interesting.
\end{quote}
This data has taken half a century to arrive. In this article we describe a crucial consistency test that has been carried out of this foundational assumption of the standard cosmological model --- which it appears to \emph{fail}.

\section{\texorpdfstring{The $\Lambda$}{\Lambda}CDM Model} 
\label{sec:LCDM}

\subsection{Its basis as an FLRW cosmology}
\label{sec:LCDM:Basis}

The assumptions of homogeneity and isotropy embodied in the Cosmological Principle considerably simplify the mathematical description of the world model, since all hypersurfaces with constant cosmic standard time are then maximally symmetric subspaces of the whole of space-time, and all cosmic tensors (such as the metric $g_{\mu\nu}$ or energy-momentum $T_{\mu\nu}$) are form-invariant with respect to the isometries of these surfaces \cite[see:][]{Weinberg:1972kfs}. These assumed symmetries thus enable a simple description of the Universe and of its dynamical evolution. The complexities of general relativity allow for many other possibilities \cite[see:][]{Krasinski:1997yxj} but these have not been much considered since the simplest solution discussed below has proved to be adequate to describe our Universe --- at least until now. Moreover it is hard to obtain exact solutions of Einstein's equations, and any departure from homogeneity and isotropy leads to rapid proliferation of the number of variables/observables.

For such a homogeneous and isotropic evolving space-time, we may choose comoving spherical coordinates (i.e. constant for an observer expanding with the universe) in which the proper interval between two space-time events is given by the Friedmann-Lema\^{ii}tre-Robertson-Walker ({\rm FLRW}) metric:
\begin{align} 
 \mathrm{d} s^{2} &= g_{\mu \nu} \mathrm{d}x^{\mu}\mathrm{d}x^{\nu} = \\
 &= \mathrm{d}t^{2} - a^{2}(t)
  \left[\frac{\mathrm{d}r^{2}}{1 - kr^{2}} + r^2(\mathrm{d}\theta^2 + \sin^2\theta \mathrm{d}\phi^2) \right] .\nonumber
\label{eq:FLRW}
\end{align}
Here $k$ is the 3-space curvature signature which is conventionally scaled (by transforming the comoving coordinate $r\to\vert{k}\vert^{1/2}r$ and $a\to\vert{k}\vert^{-1/2}a$) to be $-1$, $0$ or $+1$ corresponding to an elliptic, euclidean or hyperbolic space. The cosmic scale-factor $a(t)$ evolving in time describes the expansion of the universe; light which has been emitted in the past and is observed today at $t_0$ (when the scale-factor is $a_0$) has its wavelength redshifted by a factor $z = a_0/a -1$. 

The energy-momentum tensor then has to be of the `perfect fluid' form
\begin{align}
 T_{\mu \nu} = p g_{\mu \nu} + (p + \rho) u_{\mu} u_{\nu}\ ,
\end{align}
in terms of the pressure $p$, energy density $\rho$ and 4-velocity $u_{\mu}\equiv{\mathrm{d}}x_{\mu}/{\mathrm{d}}s$ of a comoving fluid element. 
The Einstein field equations relate $T_{\mu\nu}$ to the space-time curvature $R_{\lambda\mu\nu\kappa}$:
\begin{align}
\label{einstein}
 R_{\mu \nu} - \frac{1}{2} g_{\mu\nu} R_{c} = \frac{8\pi
  T_{\mu\nu}}{M_\mathrm{Pl}^{2}}\ ,
\end{align}
where $R_{\mu\nu}{\equiv}g^{\lambda\kappa}R_{\lambda\mu\kappa\nu}$ is the Ricci tensor, $R_{c} \equiv g^{\mu\nu} R_{\mu\nu}$ is the curvature scalar; and $M_\mathrm{Pl} \equiv (8\pi G_\mathrm{N})^{-1/2} \simeq 1.2\times10^{19}$~GeV is the Planck mass. For the FLRW metric, these equations simplify vastly to yield the Friedmann-Lema\^{\i}tre (FL) equations, one for the expansion rate $H$, also called the Hubble parameter,
\begin{align}
\label{fried1}
 H^{2} \equiv \left(\frac{\dot{a}}{a} \right)^{2} = \frac{8 \pi \rho}{3 M_\mathrm{Pl}^2} - \frac{k}{a^2}\ ,
\end{align}
as well as one for the acceleration:
\begin{align} 
\label{fried2}
 \ddot a = - \frac{4 \pi \rho}{3 M_\mathrm{Pl}^2} (\rho + 3 p) a\ .
\end{align}
Further, the conservation of energy-momentum
\begin{align} 
\label{conserv1}
 T^{\mu \nu}_{\phantom{\mu\nu};\nu} = 0\ ,
\end{align}
implies
\begin{align}
\label{conserv2}
 \frac{\mathrm{d} (\rho a^{3})}{\mathrm{d} a} = - 3 p a^{2} .
\end{align}
This can also be derived from Eqs.~(\ref{fried1}) and (\ref{fried2}) since Eqs.~(\ref{einstein}) and (\ref{conserv1}) are related by the Bianchi identities:
\begin{align}
 \left(R^{\mu \nu} - \frac{1}{2} g^{\mu \nu} R_{c}\right)_{;\mu} = 0\ .
\end{align}

In principle, a Cosmological Constant, $\lambda g_{\mu\nu}$, may be added to the field equation~(\ref{einstein}) reflecting its invariance under general local coordinate transformations.
This is equivalent to the freedom granted by the conservation equation~(\ref{conserv1}) to scale $T_{\mu\nu} {\to} T_{\mu\nu} +\Lambda g_{\mu \nu}$, so that $\Lambda$ can be related to the energy-density of the quantum vacuum \cite[see:][]{Weinberg:1988cp}:
\begin{align}
\label{Lambdavacuum}
 \langle 0 \mid T_{\mu \nu} \mid 0 \rangle = - \rho_\mathrm{vac} \, g_{\mu \nu}\ , 
 \qquad \Lambda = 8 \pi G_\mathrm{N}\rho_\mathrm{vac}\ .
\end{align}
The \emph{effective} Cosmological Constant, which would appear as an additive term $\Lambda/3$ on the rhs of the F-L equations~(\ref{fried1}) and ~(\ref{fried2}), is then the sum of two terms, which are, in a general background, quite unrelated:  
\begin{align}
\label{Lambda}
\Lambda = \lambda + 8\pi G_\mathrm{N} \rho_\mathrm{vac} .
\end{align}
All locally inertial observers must see the same quantum vacuum state, hence the equation of state of $\Lambda$ is $p=-\rho$. 

With the `equation of state' parameter $w \equiv p/\rho$ for all components, the evolution history can be constructed. For non-relativistic matter with $w \simeq 0$, $\rho_\text{m} \propto (1 + z)^{-3}$, while for radiation which has $w = 1/3$, $\rho_\mathrm{r} \propto (1 + z)^{-4}$, but for the Cosmological Constant, $w = -1$ and $\rho_\mathrm{vac} = $constant. Thus radiation was dynamically important only in the early universe (for $z \gtrsim 10^4$) and for most of the expansion history only non-relativistic matter is relevant. 
At late times, the FL equation~(\ref{fried1}) can then be rewritten as:
\begin{align}
H^2 = H_0^2\, [\Omega_\mathrm{m}(1+z)^3 + \Omega_k(1+z)^2 + \Omega_\Lambda],
\end{align}
where $\Omega_\mathrm{m} \equiv \rho_m/(3H_0^2/8\pi G_\mathrm{N})$, $\Omega_k \equiv -k/H_0^2 a_0^2$ and $\Omega_\Lambda \equiv \Lambda/3H_0^2$. Thus at the present epoch, we have the simple `cosmic sum rule':
\begin{align}
 \Omega_\mathrm{m} + \Omega_k + \Omega_\Lambda = 1,
\label{sumrule}
\end{align}
which encapsulates the FLRW models, enabling them to all be displayed on the `cosmic triangle' whose sides are the 3 parameters above \cite{Bahcall:1999xn}. Note that this can  only be done by assuming isotropy and homogeneity --- otherwise there will be additional terms and the above sum rule will no longer hold.

With its negative pressure, $\Lambda$ can in principle overcome the deceleration of the  expansion rate due to gravitating matter and reverse the sign on the rhs of Eq.~(\ref{fried2}). Such  accelerated expansion was inferred in the late 1990s from the Hubble diagram of Type~Ia supernovae which indicated: $0.8\,\Omega_\mathrm{m} - 0.6\,\Omega_\Lambda = -0.2 \pm 0.1$ \cite{SupernovaCosmologyProject:1998vns,SupernovaSearchTeam:1998fmf}. Meanwhile, temperature fluctuations had been detected in the CMB and their typical angular scale of $\sim 1^\circ$ indicated that $\Omega_k \simeq 0$ i.e.~the universe is close to being spatially flat \cite{Boomerang:2000efg}. Moreover, observations of massive clusters of galaxies indicated that $\Omega_\mathrm{m} = 0.3 \pm 0.1$ \cite{Bahcall:1998ur}. Putting all this together in the FLRW framework (\ref{sumrule}) then implies $\Omega_\Lambda \simeq 0.7$, $\Omega_\mathrm{m} \simeq 0.3$, $\Omega_\kappa \simeq
0$ \cite[see:][]{Sahni:1999gb,Peebles:2002gy}. In the subsequent two decades this conclusion has been strengthened with further observations of galaxy clustering, baryon acoustic oscillations, weak lensing, redshift space distortions etc \cite[see:][]{Frieman:2008sn,Weinberg:2013agg,Huterer:2017buf}. Combined fits to the latest datasets indicate that for the standard flat Lambda-Cold-Dark-Matter ($\Lambda$CDM) model: $\Omega_\mathrm{m}=0.315$ and $\Omega_\Lambda=0.685$ \cite[see:][]{ParticleDataGroup:2024cfk}. 

Although successful in matching a wide body of observations, the dominant component of the $\Lambda$CDM model has \emph{no} physical explanation \cite[see:][]{Sarkar:2007cx,Burgess:2025vxs}. The inferred value of the Cosmological Constant is $\Lambda \simeq 2H_0^2$ where $H_0 \equiv 100 h$~km~s$^{-1}$~Mpc$^{-1}$ with $h \simeq 0.7$, i.e. $H_0 \sim 10^{-42}$~GeV, hence $\rho_\Lambda = \Lambda/8\pi G_\mathrm{N} \sim (10^{-12} \mathrm{GeV})^4 \sim 10^{-122}M_\mathrm{Pl}^4$. However the present quantum vacuum is known to violate the $SU(2)_{\rm L}{\otimes}U(1)_{Y}$ symmetry of the electroweak interaction in the Standard Model quantum field theory.
This symmetry would have been restored at sufficiently high temperatures of ${\cal O}(10^3)$~GeV in the early universe and the energy density of the symmetric vacuum is at least a factor of $\sim 10^{60}$ higher than the maximum value of $\Lambda$ that would have enabled the Universe to expand to a size $\sim H_0^{-1}$ today. This is the as yet unsolved Cosmological Constant problem. It is further exacerbated if rather then being zero due an exact cancellation in Eq.(\ref{Lambda}), it has in fact a value of ${\cal O}(H_0^2)$ which raises the further conundrum of why it has come to dominate \emph{today}. It is worth noting in this context that it is just the assumption of the CP and the resultant sum rule (\ref{sumrule}) that effectively forces $\Lambda$ to take on such a value when inferred from cosmological data.

\subsection{Perturbations and convergence to homogeneity and isotropy}
\label{sec:LCDM:Convergence}

The smooth background defined by the FLRW metric~(\ref{eq:FLRW}) provides the underlying basis for $\Lambda$CDM cosmology. For the model to accommodate the structure that we see in our Universe, it must be endowed with perturbations, both to $g_{\mu\nu}$ and to the energy-momentum $T_{\mu\nu}$. These are required to have initially a nearly scale-free power spectrum \cite{Harrison:1969fb,Zeldovich:1972zz}, which is generally believed to have arisen from quantum fluctuations of a slowly evolving scalar field which dominated in the early universe, driving a period of inflation \cite[see:][]{Riotto:2002yw}. Coupled through the Einstein field equations~(\ref{einstein}), these perturbations, which are small $\mathcal{O}(10^{-5})$ uncorrelated Gaussian fluctuations at the time of decoupling, evolve as the universe expands and begin to grow on scales that are small enough to have entered the particle horizon. Irrespective of their evolution, they are expected to obey \emph{statistically} the same properties as their background, viz. spatial isotropy and homogeneity. This means that the underlying distributions of all their statistical descriptors are independent of location and orientation, however only for a comoving observer.

In the late universe, the perturbations have grown under gravitational instability to create $\mathcal{O}(1)$ fluctuations in the matter field that we see today as the `cosmic web'. Along with gravitational collapse came gravitational dynamics, exhibited as so-called `peculiar velocities', i.e.~velocities in addition to the Hubble recession, of galaxies and other tracers of the matter field. Therefore, as observers we are unlikely to be at rest with respect to the `cosmic reference frame' (CRF) set by the FLRW metric; i.e. we are \emph{not} comoving observers. Viewing the perturbations from this vantage point brings with it an explicit violation of isotropy -- a dipolar (to 1st order) modulation of their underlying statistical distributions, aligned with the direction of our motion, and described by a (special) relativistic boost. Knowing our relative motion, one can undo the inherent anisotropy, by de-boosting all observables accordingly, and restore the expectation of isotropy on the sky. We must therefore measure our motion with respect to the `cosmic rest frame' (often called the `CMB frame'). We describe such measurements in the next subsection, after first discussing the scale of homogeneity, the dipole anisotropy in the CMB induced by our local peculiar motion, and whether the cosmological perturbations themselves are inferred to adhere to the CP.

Statistical analyses of the large scale structure (LSS) traced by galaxies can test its adherence to the CP. Of particular interest is the scaling of galaxy counts in spheres of increasing radius $r$, that, for a homogeneous point process, should go as $r^3$ on sufficiently large scales. The scaling was in fact found to be $\propto r^2$ indicating a fractal distribution up to tens of Mpc \cite[see:][]{Coleman:1992cm}, but observed counts in the deep SDSS \cite{Hogg:2004vw} and WiggleZ \cite{Scrimgeour:2012wt} surveys are said to have demonstrated homogeneity on scales above $r \sim 70h^{-1}$~Mpc. However there are concerns because such surveys have not been large enough to provide an unbiased result, e.g.~in the WiggleZ Survey which measured redshifts for only 200,000 galaxies, the largest spheres in which galaxies were being counted and said to be scaling as $r^3$ extended in fact \emph{beyond} the survey volume --- so the numbers in the unobserved regions were simulated assuming a random (i.e.~statistically homogeneous) distribution. Such circular reasoning has been adopted as a ``bias correction'' in subsequent work which used SDSS quasars, in order to claim agreement with the expectation for the homogeneity scale in $\Lambda$CDM \citep[e.g.][]{Goncalves:2020erb}. Much bigger forthcoming surveys are required to definitively settle the issue.

The gravitational attraction of the Virgo cluster, at about $16.5$~Mpc, has long been known to produce a so-called `Virgocentric infall' in our neighbourhood. \citet{Stewart:1967ve} observed that ``If the microwave blackbody radiation is both cosmological and isotropic, it will only be isotropic to an observer who is at rest in the rest frame of distant matter which last scattered the radiation'' and thus \emph{predicted} using the velocity data in the Virgocentric infall region that there must be a dipole anisotropy in the CMB. \citet{Peebles:1968zz} calculated that an inertial observer moving with velocity $\beta=v/c$ in an isotropic blackbody radiation bath of temperature $T_0$ will measure an effective temperature that varies with angle $\theta$ with respect to the direction of motion, as:
\begin{align}
T (\theta) = T_0 \frac{\sqrt{1-\beta^2}}{(1-\beta\cos\theta)}.
\end{align}
Since our peculiar velocity is a few hundred km\,s$^{-1}$, the amplitude of the dipole in the CMB temperature should then be (expanding to 1st-order in $\beta$): 
\begin{align}
{\mathcal D}_\mathrm{CMB} = \beta \simeq 10^{-3}.
\label{eq:cmbdipole}
\end{align}
The predicted dipole anisotropy was detected subsequently by \citet{1969Natur.222..971C} \cite[for a review of such experiments, see:][]{Smoot:2007zz}. The latest measurement by the Planck satellite \cite{Planck:2018nkj} yields an inferred velocity of $369.82 \pm 0.11$~km\,s$^{-1}$ towards Galactic co-ordinates $l = (264.021 \pm 0.011)^\circ, b = (48.243 \pm 0.005)^\circ$. The spectrum of the dipole component was measured by the FIRAS instrument on the COBE satellite and is  consistent with the expected \emph{derivative} of a Planck spectrum, with an amplitude of $3.343 \pm 0.016$~mK (95\% c.l.) and a temperature of $2.714 + 0.022$~K (95\% c.l.) \cite{Fixsen:1993rd}. This is consistent with the  precisely measured $T_0=2.72548 \pm 0.00057$~K temperature of the CMB monopole \cite{Fixsen:2009ug}, lending credence to the kinematic interpretation of the dipole. 

Small-scale anisotropies are also observed in the CMB with amplitude of $\mathcal{O}(10^{-5})$; statistical analyses indicate that these are  consistent with having been drawn from an isotropic distribution~\citep{Planck:2019evm}, although the confidence with which this can be stated depends on the scales investigated, Measurements on large scales (low multipoles $\ell$) carry larger uncertainties because the $2\ell + 1$ independent samples available are fewer; this is the so-called `cosmic variance'. Even so there are indications of deviations from statistical isotropy, collectively referred to as low-$\ell$ anomalies, that are in mild ($\sim1-3\sigma$) tension with expectations~\cite[see:][]{Schwarz:2015cma}. The origin of these anomalies has so far not been established, nor is it clear whether these are statistical flukes, systematic errors or true deviations from statistical isotropy. Either way, these anisotropies are small, constraining the size of observable deviations from the CP at $z \simeq 10^3$.

Assuming statistical isotropy of the CMB fluctuations in the CMB frame, the compression of the angular scale --- and consequent shift in the CMB power spectrum --- in the direction of motion, yields an estimate of our velocity with respect to the CMB frame that can be checked against the purely kinematic interpretation of the CMB dipole \citep{Challinor:2002zh,Burles:2006xf}. The first such measurement \cite{Planck:2013kqc} found $v=384\pm78 (\mathrm{stat.})\pm115 (\mathrm{syst.})$~km~s$^{-1}$ towards the CMB dipole hotspot. Although this still allows up to about 40\% of the CMB dipole to be non-kinematic in origin ~\citep{Schwarz:2015cma}, a subsequent Bayesian analysis by \citet{Saha:2021bay} \citep[see also:][]{Ferreira:2020aqa} detected the expected standard kinematic signal at $\sim4.5\sigma$. Planck also measured the dipole modulation of the thermal Sunyaev-Zeldovich effect, again finding consistency with the kinematic interpretation  \citep{Planck:2020qil}, however this effect is agnostic towards whether the CMB dipole is kinematic in origin or intrinsic \cite{Notari:2015daa}.

In any case, if the CP is to hold, the rest frame of the CMB should coincide with the rest frame of distant galaxies. In other words, the mean motion of galaxies (w.r.t.~CMB rest frame) inside a sphere of radius $r$ (in comoving coordinates), must be a decreasing function of~$r$. Observation of galaxies, clusters and superclusters in the local Universe show, however, a highly anisotropic environment out of large distances. The Milky Way along with the Local Group of neighbouring galaxies, e.g. Andromeda, all participate in a coherent `bulk flow' towards a direction close to that of the CMB dipole hotspot. Although peculiar velocities are hard to measure at high redshift, nearby we have a good estimates of them via independent measures of distance. If the Universe becomes isotropic and homogeneous on large scales  and the rest frame of matter is the same as the rest frame of the CMB, then the bulk flow should die out on a scale exceeding $\sim150h^{-1}$~Mpc where the Universe is said to become sensibly homogeneous \cite{Hogg:2004vw, Scrimgeour:2015khj, Goncalves:2020erb}.

It was realised that one has to go well beyond the Virgo cluster, to at least the Hydra-Centaurus supercluster (known today as the `Great Attractor'), to account for the CMB dipole \cite{1985ApJ...294...81T}. For some  years, the CMB rest frame was assumed to be at the scale of the Great Attractor, i.e.~at about $50h^{-1}$~Mpc. However the work of the `Seven Samurai' \cite{Dressler:1986rv} marked a turning point by firmly establishing that the CMB dipole, if kinematic in origin and due to the motion of the Local group, \emph{cannot} be primarily the result of gravitational attraction by structures inside the boundary of the Great Attractor. Subsequent studies up until now have shown that even up to the Shapley concentration (at around $120h^{-1}$~Mpc) this bulk flow continues unabated \citep[see {\it e.g.}:][]{Lavaux:2008th,Colin:2010ds,Feindt:2013pma}. The most recent studies using the CosmicFlows-4 survey have established that the bulk flow continues well beyond Shapley \citep{Watkins:2023rll}. This is in conflict with the expectation in the $\Lambda$CDM model at $\sim3-5\sigma$ \cite{Watkins:2023rll,Whitford:2023oww,Duangchan:2025uzj}.

Nevertheless, there is little room to interpret the CMB dipole as anything other than a kinematic effect induced by our peculiar motion with respect to the CMB frame, a frame that must be shared by LSS. This is what makes the test we discuss below by \citet{1984MNRAS.206..377E} so powerful: comparing the dipole apparent in the LSS at $z\sim1$ with the kinematic prediction from the $z\sim1100$ CMB dipole is a critical consistency test of the foundational assumption, inherent in all FLRW-based cosmologies such as $\Lambda$CDM, that the evolution of the universe on cosmological scales is described by a direction-\emph{independent} scale factor $a(t)$ and that the LSS, arising from the $\mathcal{O}(10^{-5})$ inhomogeneities imprinted on the CMB, must therefore share the same frame. 
If the \citeauthor{1984MNRAS.206..377E} test does not confirm that the dipole in cosmologically distant sources is consistent with a local boost of $\sim370$~km~s$^{-1}$ (as inferred from the CMB dipole) then the assumption of homogeneity and isotropy on large scales is observationally contradicted.

\section{The Ellis \& Baldwin Test}
\label{sec:EBtest}

The key insight of \citet{1984MNRAS.206..377E} was that cosmologically distant sources should exhibit the same dipole anisotropy due to our local motion as is evident in the CMB. They noted that this directly tests whether the `rest frame' of distant matter is indeed the same as that of the CMB, as is expected in a FLRW universe. 

We now discuss how the cosmic matter dipole arises, first in the language of special relativity used by \citeauthor{1984MNRAS.206..377E}, and then including general relativistic effects.

\subsection{Derivation in special relativity}
\label{sec:EBtest:SR}

The prescription of \citet{1984MNRAS.206..377E} relies entirely on observed quantities: starting with a catalog of extragalactic sources with observed spectral flux density $S(\nu)$ at frequency $\nu$, a flux-limited sample is selected by removing sources with $S<S_*$, and its projected number density ${\rm d}N/{\rm d}\Omega$ is investigated. If we are moving with respect to the frame in which the sources have projected number density ${\rm d}N/{\rm d}\Omega_0$, a Lorentz boost translates from that frame to ours. With the flux density cut, two distinct effects determine the number density observed in our frame: Doppler boosting of frequencies and flux densities, and angular aberration of the source positions on the sky with respect to our direction of motion.\footnote{While at this point, most discussions start making assumptions regarding source spectra $S(\nu)$, distributions of number counts ${\rm d}N/{\rm d}\Omega$, redshift-independence, etc~\citep[e.g.,][]{Rubart:2013tx, Itoh:2009vc, Tiwari:2013vff}, we prefer to retain the general form just a bit longer before we outline the considerations leading to the \citeauthor{1984MNRAS.206..377E} formula (\ref{eq:Dkin}).}

First consider the Lorentz boost of the photon frequency when our velocity is $\beta \equiv v/c$:
\begin{align}
    \nu &= \gamma(1+\beta\cos\theta)\,\nu_0 
    \nonumber\\
    &\approx (1+\beta\cos\theta)\,\nu_0 \equiv \eta\,\nu_0,
\label{eq:nu}
\end{align}
where $\theta$ is the angle between the direction of observation and that of the boost, and $\gamma=(1-\beta^2)^{-1/2}$ is the Lorentz boost factor. Using this and photon number conservation leads to the Lorentz invariant for the source spectrum:
\begin{align}
    \frac{S(\nu)}{\nu} = \frac{S_0(\nu_0)}{\nu_0},
\label{LI}
\end{align} 
Relativistic aberration of the unit solid angle gives:
\begin{align}
    {\rm d}\Omega &= (1+\beta\cos\theta)^{-2}{\rm d}\Omega_0 = \eta^{-2}{\rm d}\Omega_0 \nonumber \\
    &\approx (1-2\beta\cos\theta){\rm d}\Omega_0.
\label{eq:dOmega}
\end{align}
Our aim is to find an expression for the projected integral source counts per unit solid angle ${\rm d}N/{\rm d}\Omega(S>S_*)$ in terms of their rest-frame distribution above the same threshold: $S_*$, ${\rm d}N/{\rm d}\Omega_0(S>S_*)$. For this it is important to model the functional form in the \emph{immediate} vicinity of $S_*$. We can locally approximate the integral source counts as a power law, ${\rm d}N/{\rm d}\Omega(S>S_*)\propto S_*^{-x(S_*)}$ with index $x$, such that:
\begin{align}
    x=x(S_*) \equiv -\frac{\partial}{\partial \ln S_*}\left[\ln \frac{{\rm d}N}{{\rm d}\Omega}(S>S_*)\right].\label{eq:x}
\end{align}
\citet{1984MNRAS.206..377E} discussed a constant, i.e. flux-independent, power-law index $x$. This assumption need only hold locally around $S_*$, although this has not always been appreciated. There have been attempts to fit the observed projected integral source counts with a more flexible (e.g. log-normal) function over a wide range of flux densities $S$ \citep{Tiwari:2013vff,Siewert:2020krp}. However this can actually 
mispredict the kinematic matter dipole, 
since sources far from the threshold $S_*$ play no role in determining the dipole amplitude \cite{vonHausegger:2024jan}, as explained in \S~\ref{sec:EBtest:Practice}.

Similarly, the source spectrum can be modelled as a power law, $S(\nu)\propto\nu^{-\alpha}$, so Lorentz invariance yields
\begin{align}
    S(\nu) = \eta^{1+\alpha}S_0(\nu) \approx (1+(1+\alpha)\beta\cos\theta)S_0(\nu).
\end{align}
The power-law assumption is appropriate for a wide range of source types, such as radio and mid-infrared active galactic nuclei (AGN) with which this test has been performed so far \citep[e.g.,][]{Secrest:2022uvx}, however care is needed if one is to perform similar studies with sources whose spectra have sharp spectral features. Denoting by $\alpha$ the average spectral index of sources in the vicinity of $S_*$, we find:
\begin{align}
    \frac{{\rm d}N}{{\rm d}\Omega}&(S>S_*) = \eta^{2+x(1+\alpha)}\frac{{\rm d}N}{{\rm d}\Omega_0}(S>S_*) \nonumber\\
    &\approx \left[1+(2+x(1+\alpha))\beta\cos\theta\right]\frac{{\rm d}N}{{\rm d}\Omega_0}(S>S_*),
\label{eq:dNdO}
\end{align}
where the second line is a linear approximation in observer velocity $\beta$. The angular dependence of the second term constitutes a dipolar modulation of the integral source counts on the sky. It is called the `kinematic matter dipole' and has amplitude \cite{1984MNRAS.206..377E}:
\begin{align}
    \mathcal{D}_{\rm kin} = \left[2+x(1+\alpha)\right]\beta ,
\label{eq:Dkin}
\end{align}
so is larger than the CMB dipole amplitude (\ref{eq:cmbdipole}) by a factor of $2+x(1+\alpha)$.

\subsection{General relativistic corrections and redshift dependence}
\label{sec:EBtest:GR}

The derivation above of the kinematic matter dipole is in terms of quantities observed in a projected sample on the sky. However, source catalogs now also contain information on their redshifts so a better understanding of the kinematic matter dipole expectation is warranted in view of the possible variation of the key observables along our past lightcone. Following \citet{Maartens:2017qoa}, we repeat the special relativistic derivation considering source redshifts before introducing an important distinction between observed redshifts and those tracing the background metric. The perturbations to background quantities considered here involve only those induced by the boost. For a full treatment of all first-order perturbations see \citet{Domenech:2022mvt} who obtain the same final expression.\\

First recall the Doppler shift of redshifts
\begin{align}
    1+z &= (1-\beta\cos\theta)(1+z_0), \nonumber\\
    {\rm d}z &= (1-\beta\cos\theta){\rm d}z_0,\nonumber\\
    z_0-z &= \beta\cos\theta(1+z_{0}),
\label{eq:dz}
\end{align}
along with the aberration effect on the solid angle~(\ref{eq:dOmega}).

Let us express the source counts per redshift interval and unit solid angle in terms of their rest-frame density using number count conservation, and then expand around their observed redshifts $z$ as:
\begin{align}
    \frac{{\rm d}N}{{\rm d}\Omega{\rm d}z}(z) =& \frac{{\rm d}N}{{\rm d}\Omega_0{\rm d}z_0}(z_0)\frac{{\rm d}\Omega_0{\rm d}z_0}{{\rm d}\Omega{\rm d}z} \nonumber \\
    \approx& \frac{{\rm d}N}{{\rm d}\Omega_0{\rm d}z_0}(z_0)\left(1+3\beta\cos\theta\right) \nonumber \\
    \approx& \left(1+3\beta\cos\theta\right)\times\left[\frac{{\rm d}N}{{\rm d}\Omega_0{\rm d}z_0}(z) + \right.\\
    &\quad\qquad\quad+\left.\frac{{\rm d}}{{\rm d}z_0}\frac{{\rm d}N}{{\rm d}\Omega_0{\rm d}z_0}(z)(z_0-z)\right]\nonumber\\
    \approx&\frac{{\rm d}N}{{\rm d}\Omega_0{\rm d}z_0}(z)\left(1+3\beta\cos\theta\right)+\\
    &\quad\qquad\quad+\frac{{\rm d}}{{\rm d}\log(1+z)}\frac{{\rm d}N}{{\rm d}\Omega_0{\rm d}z_0}(z)\beta\cos\theta,\nonumber
\end{align}
where we used Eqs.~(\ref{eq:dOmega}) and (\ref{eq:dz}), and further dropped the explicit dependencies on direction $\hat{\bf n}$, or $\hat{\bf n}_0$ as the differences are 2nd-order in $\beta$. The last line is obtained by further approximating the last term to be linear in $\beta$. Collecting terms yields:
\begin{align}
    \frac{{\rm d}N}{{\rm d}\Omega{\rm d}z}(z) &\equiv \frac{{\rm d} N}{{\rm d}\Omega_0{\rm d}z_0}(z)\left[1+\mathcal{D}_{\rm kin}(z)\cos\theta\right],
\end{align}
where the redshift dipole amplitude for a catalog \emph{without} a flux limit is defined as
\begin{align}
    \mathcal{D}_{\rm kin}(z) \equiv \left[3+\frac{{\rm d}}{{\rm d}\log(1+z)}\log\left(\frac{{\rm d}N}{{\rm d}\Omega_0{\rm d}z_0}(z)\right)\right]\beta.
\end{align}
Expressed in terms of the comoving number density $n(z)\equiv(1+z)^3Hr^{-2}{\rm d}N/{\rm d}\Omega_0{\rm d}z_0(z)$, where $r$ is the comoving distance, gives
\begin{align}
    \mathcal{D}_{\rm kin}(z) = \left[3+\frac{\dot{H}}{H^2}+\frac{2(1+z)}{rH}-b_e(z)\right]\beta,\label{eq:Dkinz1}
\end{align}
where $b_\text{e}(z)\equiv-{\rm d}\log((1+z)^{-3}n(z))/{\rm d}\log(1+z)$ is the `evolution bias' of the source population and $\dot{H}={\rm d}H/{\rm d}t$ is the derivative of the Hubble parameter $H$ with respect to time. Here we use ${\rm d}r/{\rm d}z=H^{-1}$ and ${\rm d}H/{\rm d}z=-\dot{H}H^{-1}(1+z)^{-1}$. Given the approximations that are appropriate here, we can think of $z_0$ as the background, i.e.~true cosmological redshift $\overline{z}$. The distribution ${\rm d}N/{\rm d}\Omega_0{\rm d}z_0(z)$ is moreover independent of direction. We see that the amplitude of the dipolar modulation per redshift interval picks up additional contributions from both volume perturbations (2nd and 3rd terms in Eq.~(\ref{eq:Dkinz1})), and from the evolution of the source counts with redshift (4th term in Eq.~(\ref{eq:Dkinz1})). The latter arises because of the boosted observer probing source counts at different background redshifts depending on direction.\\

However, we are interested in source counts integrated above some flux density limit, viz. the integral source counts per redshift and unit solid angle. This can also be expressed in terms of source luminosity~\citep{Dalang:2021ruy}
\begin{align}
    \frac{{\rm d}N}{{\rm d}\Omega{\rm d}z}(z,S>S_*) = \int_{S_*}^\infty{\rm d}S\frac{{\rm d}N}{{\rm d}\Omega{\rm d}z{\rm d}S}(z,S) \nonumber\\
    = \int_{L_*}^\infty{\rm d}L \frac{{\rm d}N}{{\rm d}\Omega{\rm d}z{\rm d}L}(z,L) = \frac{{\rm d}N}{{\rm d}\Omega{\rm d}z}(z,L>L_*(z)),
    \label{full}
\end{align}
where $L$ denotes the source luminosity, whose threshold $L_*=L_*(z)$ depends on redshift (and therewith direction) for $S_*$ held constant~\citep{Alonso:2015uua}.\footnote{$L_*$ and $S_*$ further differ in their assigned frequency which however does not impact the expansion below as the corresponding terms are varied with the redshift held constant, which also keeps the relation between source and observed redshift invariant.} The derivation now proceeds similarly with a few re-definitions and considering angular variations in luminosity around its average (viz. the source luminosity $L_{*,0}$ as inferred by an unboosted observer), in addition to those in redshift as seen above:
\begin{align}
    \frac{{\rm d}N}{{\rm d}\Omega{\rm d}z}(z,&S>S_*) = \frac{{\rm d}N}{{\rm d}\Omega_0{\rm d}z_0}(z_0,L>L_*)\frac{{\rm d}\Omega_0{\rm d}z_0}{{\rm d}\Omega{\rm d}z} \nonumber\\
    \approx& \frac{{\rm d}N}{{\rm d}\Omega_0{\rm d}z_0}(z_0,L>L_*)(1+3\beta\cos\theta) \nonumber\\
    \approx& (1+3\beta\cos\theta)\times\left[\frac{{\rm d}N}{{\rm d}\Omega_0{\rm d}z_0}(z,L>L_{*,0})+\right. \nonumber\\
    &\qquad+\frac{\partial}{\partial z_0}\frac{{\rm d}N}{{\rm d}\Omega_0{\rm d}z_0}(z,L>L_{*,0})(z_0-z)+\nonumber\\
    &\qquad+\left.\frac{\partial}{\partial L_{*,0}}\frac{{\rm d}N}{{\rm d}\Omega_0{\rm d}z_0}(z,L>L_{*,0})(L_*-L_{*,0})\right]
\end{align}
The last term, with help of the relation
\begin{align}
    L_*-L_{*,0} &= 4\pi\frac{d_L^2-d_{L,0}^2}{(1+z)} \nonumber\\
    &\approx 2L_{*,0}\frac{d_L-d_{L,0}}{d_{L,0}} \nonumber\\
    &\approx 2L_{*,0}\frac{\beta\cos\theta}{Hr}(1+z_0),
\end{align}
yields the `magnification bias' $x(z)$, defined similarly to Eq.~(\ref{eq:x}):
\begin{align}
    x(z_0) \equiv -\frac{\partial}{\partial\log L_{*,0}}\log\left(\frac{{\rm d}N}{{\rm d}\Omega_0{\rm d}z_0}(z_0,L>L_{*,0})\right).
\end{align}
Also the `evolution bias' of this sample is 
\begin{align}
    b_e(z_0) \equiv -\frac{\partial\log[(1+z_0)^{-3}\mathcal{N}(z_0,L_{*,0})]}{\partial\log(1+z_0)},\label{eq:be2}
\end{align}
 in terms of the comoving number density $\mathcal{N}(z_0,L_{*,0})\equiv(1+z)^3Hr^{-2}{\rm d}N/{\rm d}\Omega_0{\rm d}z_0(z_0,L>L_{*,0})$. Approximating all terms to be linear in $\beta$,
this gives the kinematic redshift dipole amplitude for a flux-limited catalog to be:
\begin{align}
    \mathcal{D}_{\rm kin}(z)=\left[3+\frac{\dot H}{H^2}+\frac{2(1+z)(1-x(z))}{rH}-b_e(z)\right]\beta\label{eq:Dkinz2}
\end{align}
Written in this form \citep{Maartens:2017qoa}, the relation to the redshift-integrated kinematic matter dipole~(\ref{eq:Dkin}) is not immediately evident. This connection was made by \citet{Nadolny:2021hti} and \citet{Dalang:2021ruy},\footnote{See also \citet{Domenech:2022mvt,Lacasa:2024ybp} for complementary derivations.} where assumptions about the power-law nature of the source spectrum were imposed, viz. $L(z)=4\pi r^2 S (1+z)^{1+\alpha(z)}$. Expanding the evolution bias (\ref{eq:be2}) in terms of its total derivative and inserting into Eq.~(\ref{eq:Dkinz2}) yields
\begin{align}
    \mathcal{D}_{\rm kin}(z) = &\left[3+x(z)(1+\alpha(z))\right]\beta+\label{eq:Dkinz3}\\
    &+\frac{{\rm d}}{{\rm d}\log(1+z)}\log\left(\frac{{\rm d}N}{{\rm d}\Omega_0{\rm d}z_0}(z,S>S_*)\right)\beta.\nonumber
\end{align}
We recover a recognisable expression for the kinematic matter dipole if the normalised integral source counts,
\begin{align}
    f(z)\equiv\frac{{\rm d}N}{{\rm d}\Omega{\rm d}z}(z,S>S_*)\bigg/\int_0^\infty{\rm d}z\,\frac{{\rm d}N}{{\rm d}\Omega{\rm d}z}(z,S>S_*),
\end{align}
are assumed to vanish at the boundaries, such that the boundary term in the second line of Eq.~(\ref{eq:Dkinz3}) returns $-\beta$:
\begin{align}
    \mathcal{D}_{\rm kin} =& \int_0^\infty{\rm d}z\,f(z) \mathcal{D}_{\rm kin}(z) = \\
    =& \int_0^\infty{\rm d}z\,f(z) \left[2+x(z)(1+\alpha(z))\right]\beta\label{eq:Dkinint}
\end{align}
In practice, $x$ and $\alpha$ of Eq.~(\ref{eq:Dkin}) are computed using the projected integral source counts~(\ref{eq:dNdO}) at the flux limit as explained above. Hence if we define the \emph{average} values
\begin{align}
    x\equiv\int_0^\infty{\rm d}z\,f(z)x(z), && \alpha\equiv\int_0^\infty{\rm d}z\,\tilde f(z)\alpha(z),
\end{align}
with
\begin{align}
    \tilde f(z) \equiv \frac{{\rm d}N}{{\rm d}\Omega{\rm d}z{\rm d}S}(z,S_*)\bigg/\int_0^\infty{\rm d}z\,\frac{{\rm d}N}{{\rm d}\Omega{\rm d}z{\rm d}S}(z,S_*),
\end{align}
then Eq.~(\ref{eq:Dkinint}) reduces exactly to the form (\ref{eq:Dkin}), without explicit reference to the (usually unknown) luminosity function~\citep{vonHausegger:2024jan}. 



\subsection{Conditions for validity}
\label{sec:EBtest:Conditions}

As discussed in \S~\ref{sec:EBtest:AlphaX}, the prescription for the kinematic dipole given by \citet{1984MNRAS.206..377E} concerns a local effect: their prediction (\ref{eq:Dkin}) for $\mathcal{D}_\mathrm{kin}$ is thus \emph{unaffected} by the type, redshift, or evolutionary history of the sources that define the frame relative to which we are moving, as long as the following conditions are met:

\begin{enumerate}

\item Our velocity is small enough that $\mathcal{O}(\beta^2)$ terms in the full expression (\ref{full}) for ${\rm d}N(>S) / {\rm d}\Omega$ may be neglected. 

\item The mean source spectral index $\alpha$ is measured at the flux limit $S_*$ within the interval $\Delta S = 2(1+\alpha)\beta\,S_*$; in this interval, variations in the \textit{mean} spectral index $\delta\alpha$ must be small enough that the corresponding dipole amplitude variation $\delta\mathcal{D}_\mathrm{kin} = x\delta\alpha\,\beta$ is negligible compared with the expected amplitude: $\delta\mathcal{D}_\mathrm{kin}/\mathcal{D}_{\rm kin} \ll 1$. Since $\beta \sim 10^{-3}$ this means in practice up to $\sim1\%$ shifts in the flux limit.

\item The value of $x$ is also measured at the flux limit and any deviations are negligibly small for up to $
\sim1\%$ shifts in the flux limit (as above).

\item The \emph{observed} mean spectral energy distribution (SED) of the sources can be described by a power law within the passband that defines the flux limit (see \S~\ref{sec:EBtest:PowerLaw} for specifics).
\end{enumerate}

\subsection{Practical considerations}
\label{sec:EBtest:Practice}

\subsubsection{Determination of survey sensitivity limit}
\label{sec:EBtest:Sensitivity}

It should be noted that the quoted sensitivity limit of a catalog may have been established using small patches of sky, in which much deeper surveys have been done to estimate the reliability and completeness of the catalog. Hence, we recommend that tests of source density against possible systematics be undertaken to ensure that these do not affect the final results. 
For example, in the NVSS catalog there is a declination dependent systematic due to the Very~Large~Array D-to-DnC configuration change at the northernmost and southernmost limits of the survey. Studies of the radio galaxy dipole have generally established that the NVSS uniform sensitivity limit is between $\sim10$~and~$15$~mJy, depending on how other systematics such as the Galactic synchrotron background are controlled for. This is several times larger than the 99\% completeness at 3.4~mJy figure reported in \citet{Condon:1998iy}, probably because this figure was derived from a comparison to a deep radio survey of just a few square degrees \citep{1985A&AS...61..517K}. 

An example of where checks against \emph{a priori} systematics have not been carried out is in tests of the dipole amplitude and direction as a function of flux cut. While these tests do seek to determine the sensitivity limit of a survey, they typically do so by assuming that the dipole should converge to a particular direction as the flux cut is adjusted, usually the direction of the CMB dipole. Even if the dipole does converge to a certain direction, this does not necessarily indicate that the sensitivity limit has been found, as the type of sources selected at a given flux cut may preferentially align (or misalign) with a particular direction for astrophysical reasons. Especially in the case of the CMB dipole direction, there is no reason why a dipole inconsistent in amplitude with the kinematic prediction from the CMB should nevertheless align with it. Such assumptions can cause important new information to be overlooked, so it is prudent to focus on eliminating \emph{a priori} observational systematics and attempt to physically interpret the found dipole only after such systematics have been accounted for. 

\subsubsection{Determination of \texorpdfstring{$\alpha$}{alpha} and \texorpdfstring{$x$}{x}}
\label{sec:EBtest:AlphaX}

To recapitulate, the \citet{1984MNRAS.206..377E} test concerns a local effect, viz. the relativistic Doppler boosting and aberration of source counts in a flux-complete survey due to the observer's motion with respect to the rest frame of the sources. The apparent change in the number density of sources with the angle away from the direction of motion is a function of the power-law spectral index of the sources $\alpha$, and the rate of increase in the number of sources per unit solid angle above some flux density (at that flux density value) denoted by the power-law exponent $x$:
\begin{align}
\frac{{\rm d}N}{{\rm d}\Omega}(>S_\nu) \propto S_\nu^{-x}.
\label{eq:dndOmega}
\end{align}
\noindent 
In practice, no astrophysical source has a spectral energy distribution described at all frequencies by a single power law with index $\alpha$. Fortunately, the flux limit of a survey is typically defined at one frequency for radio wavelengths, where the SED is close to a power~law, or in one (relatively narrow) band within which the SEDs of sources may be treated as a power law. There are emission lines that fall within the observing band for quasars at certain redshifts, but the enormous range of observed quasar luminosities ($\gtrsim 4$ orders of magnitude) ensures that at a given apparent magnitude (i.e., flux limit) the aggregate or mean SED of quasars in a survey is smoothed by the corresponding wide range of redshifts and can be treated as a power law in most passbands (e.g., WISE W1 at 3.4~$\mu$m), as  demonstrated in the next section.

Second, no type of astrophysical source has a single value of $x$ at all flux levels. For example, it is well known that the flux distribution becomes steeper for radio galaxies at high flux levels \citep[see Figure~1 in][]{Rubart:2013tx}. However, the relevant value of $x$ is that very near the flux limit of the survey, because sources at fluxes brighter than the small range within which the kinematic modulation occurs are counted \emph{once and only once}, irrespective of their value of $x$. Thus, modification to the \citet{1984MNRAS.206..377E} test to account for $x$ not being a simple power~law for all fluxes is not physically justified: the dipole source count modulation due to observer motion occurs only at the flux limit of the survey. However, the modulation in source counts due to Doppler boosting will cause the observer to sample different rest-frame values of the flux density for a given observed value. If the rest-frame value of $x$ is not a single power~law for all fluxes then there will be differences in $x$ as a function of angle with respect to the observer's motion. Given that the observer's velocity is small ($\beta \ll 1$), the fractional minimum-to-maximum difference in observed flux densities relevant to any misestimation of $x$ 
is bounded by:
\begin{align}
\frac{\Delta S}{S} < 2 (1 + \alpha) \beta ,
\label{eq:deltaSoverS}
\end{align}
\noindent where the inequality reflects that most sources are not observed near the direction of our motion. 
For a typical spectral index near $\alpha=1$ and the CMB dipole-based value of $\beta = 1.23\times10^{-3}$, the fractional flux difference is $<0.5\%$, a difference much smaller than that in which deviations from a power-law parameterisation of $x$ become noticeable. Hence the kinematic prediction by \citet{1984MNRAS.206..377E} of the dipole modulation in source counts is accurate so long as $x$ is measured at the flux limit of the survey.

Third, relativistic Doppler boosting will bring sources intrinsically fainter than the survey flux limit above the detection threshold in the direction of motion, conversely lowering intrinsically brighter sources in the opposite direction below the threshold. There will therefore be intrinsic differences in which part of the source SEDs are observed due to the respective blue- and redshifts, as well as the distance to the source. However, for the small velocities ($\beta \sim 10^{-3}$) under consideration, the fractional difference in redshift corresponds to a time difference of less than $\sim10$~Myr for sources at $z\sim1$, which is quite insignificant regarding which part of the source SED is observed. A much larger ($\sim100$~times bigger) effect is the smoothing of source SEDs over the wide range of redshifts observed in quasars and radio galaxies.

\subsubsection{Power-law approximation of spectral energy distributions} 
\label{sec:EBtest:PowerLaw}

As stated in \S~\ref{sec:EBtest:Conditions}, the \citet{1984MNRAS.206..377E} is valid so long as the \emph{observed} mean SED of the sources can be described by a power law within the passband defining the flux limit. For radio sources at low frequency ($\lesssim20$~GHz) observed using a configuration sensitive to extended emission (such as the NVSS), the SEDs are dominated by optically thin synchrotron and can indeed be reliably described by a power law with spectral index of $\alpha\simeq 0.75$. 

Quasars, on the other hand, exhibit highly complex SEDs with both thermal and non-thermal continua, strong and often broadened emission features, and contributions from stellar populations of their host galaxies, so it is not \emph{a~priori} obvious why their SEDs in a given passband should be treated as power laws. Fortunately, the enormous range of redshifts observed in quasars has the effect of smearing out steeply changing or discontinuous features such as emission lines, creating a much smoother observed mean spectrum. 

\begin{figure}
\includegraphics[width=\columnwidth]{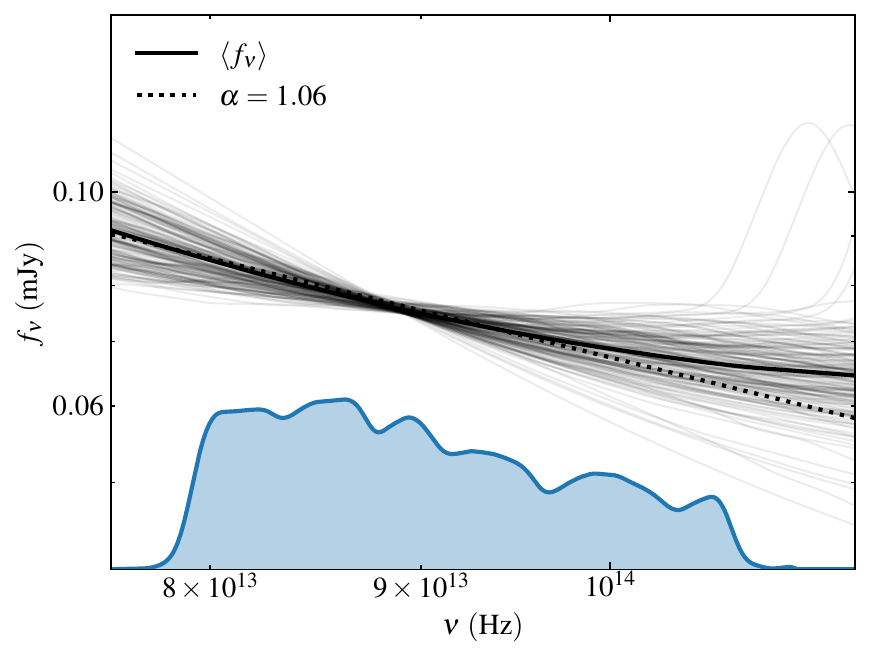}
\caption{Quasar spectral flux densities at flux limit for the sample used by \citet{Secrest:2022uvx} within the WISE W1 passband, shown at bottom for reference. The fainter lines are individual quasars while the solid black line is the mean. The dotted line shows a power law with $\alpha=1.06$ used by \citet{Secrest:2022uvx}, derived from the $W1-W2$ color. While the individual quasars have varying spectral indices and some show emission features, on average the flux density within the passband is close to a power law.}
\label{fig: mean_quasar_sed}
\end{figure}

To show this, we matched the sample of quasars from \citet{Secrest:2022uvx} to the SDSS/BOSS ``specObj'' table for DR17, using primary spectra and a match tolerance of 1$^{\prime\prime}$. After removing sources with redshift warnings or invalid measurements, as well as sources beyond $z>4$ that are likely Ly-$\alpha$ misidentifications, we produced a sample of 210,987 quasars with reliable redshifts. The redshift distribution is comparable to that found in \citet{Secrest:2020has}. We estimated the spectral energy distributions for each source by choosing the linear combination of AGN and starburst SED templates from \citet{Assef:2009wg} that is closest in $W1-W2$ color. This accounts for the increasing fraction of host galaxy emission in objects at lower redshift, although on average the AGN makes up 91\% of emission. We selected the objects at the flux limit ($W1\sim16.5$) and calculated their mean mid-IR spectrum, shown in Figure~\ref{fig: mean_quasar_sed}. As can be seen, while a few individual sources do show emission features and there is a considerable range in per-source spectral index, on average the spectrum of quasars at the flux limit is close to a power law and the power-law index of 1.06 derived by \citet{Secrest:2022uvx} using the $W1-W2$ color accurately captures the mean SED within the $W1$ band where the flux cut was made.

\subsection{The history of E \& B tests}
\label{sec:EBtest:History}

Having presented the theoretical background as well as the practical considerations that immediately follow therefrom, we briefly review the two decades of studies that have implemented the \citeauthor{1984MNRAS.206..377E} test. It is only recently that enough data has been available for statistically significant measurements of the matter dipole to be made. Apart from statistics, careful treatment of systematics and methodology is essential for the accuracy of this test.
The lessons learnt need to be retained for future while we await new data from ongoing and future experiments.\\

\begin{figure}[tbh]
\includegraphics[width=\columnwidth]{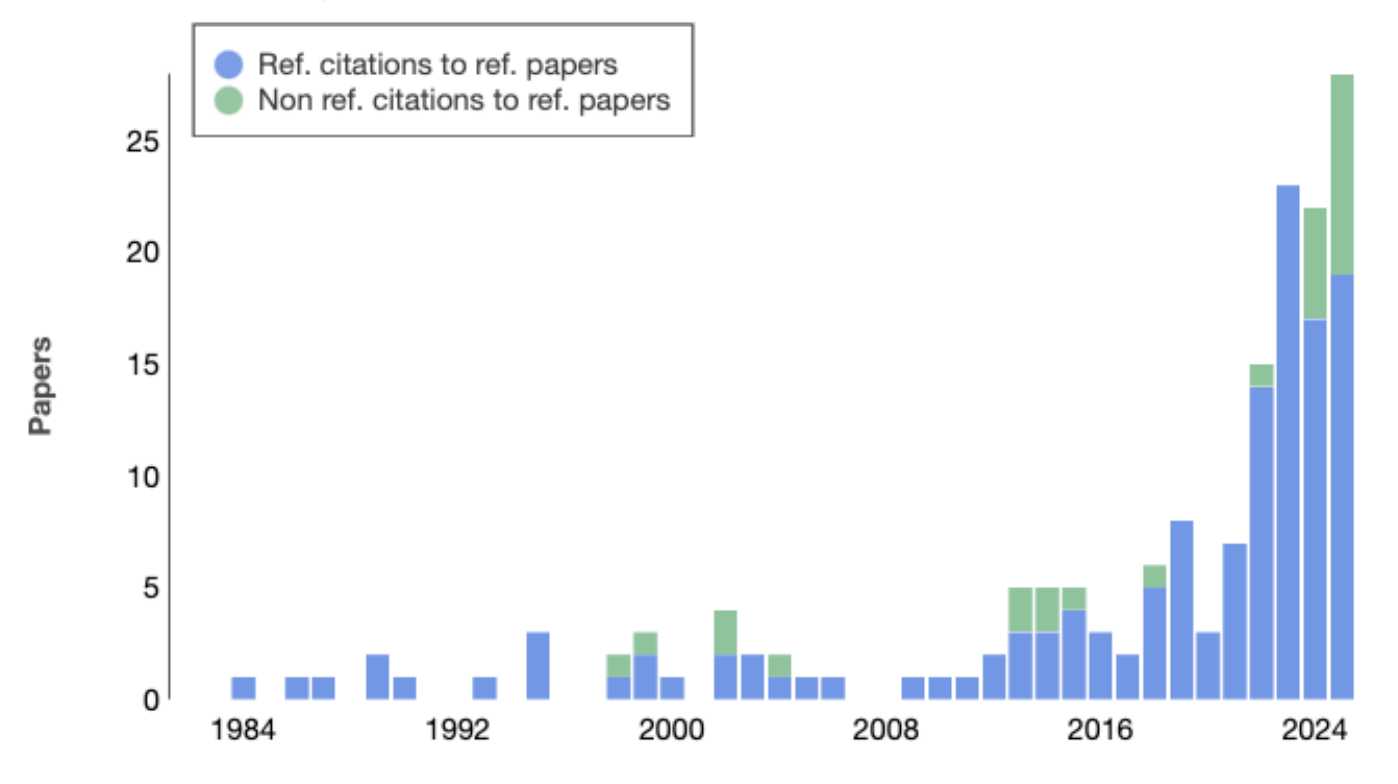}
\caption{Citations to \citet{1984MNRAS.206..377E} as of September 2025, from the \href{https://ui.adsabs.harvard.edu/abs/1984MNRAS.206..377E/abstract}{Astrophysics Data System}.}
\label{EB84_citations}
\end{figure}
As of writing, there are 163 citations to \citet{1984MNRAS.206..377E} indexed by ADS, nearly half of which have been in the last three years as seen in Figure~\ref{EB84_citations}. In the following we offer an explanation for this trend and for the corresponding rapid development of the field now known as the `Cosmic Dipole Anomaly'. A brief comment commemorating ``Forty years of the Ellis-Baldwin test'' has been presented by \citet{Secrest:2025nbt}, and a broad overview of key results by \citet{2025RSPTA.38340027S}.

\subsubsection{The first detection}
\label{sec:EBtest:History:FirstDetection}

With the release of the 1.4~GHz NRAO~VLA~Sky~Survey at the end of the 20th century \citep{Condon:1998iy}, a predominantly extragalactic catalog with sky coverage and source counts sufficient to constrain the cosmic dipole finally became available. It was thus a relief to the cosmological community that the first such test reported agreement with the 370~$\rm{km}\,\rm{s}^{-1}$ kinematic prediction from the CMB at $\sim1.5\sigma$ \citep{Blake:2002gx}, with associated commentary by \citet{2002Natur.416..132E} celebrating the result.

The measured matter dipole direction agreed well with that of the CMB dipole within uncertainties, so attention focussed on comparing the measured dipole amplitude with the predicted value~(\ref{eq:Dkin}). Despite the initial agreement, later publications have generally found that the dipole amplitude in the NVSS is \emph{not} consistent with the kinematic prediction \citep[e.g.][]{Singal:2011dy,Rubart:2013tx,Tiwari:2015tba}, primarily due to how the significance of the result was being determined. While \citet{Blake:2002gx} stated that their result was only $1.5\sigma$ from the kinematic expectation, this number is an average of the per-flux density cut values in their Table~1 down to 15~mJy, below which they find that a dipole model is a poor fit to the data. Such an average is not well-motivated, among other reasons, due to the fact that subsamples with lower flux cuts contain all sources of those subsamples with higher flux cuts, rendering the various samples statistically dependent. A more principled approach would be to quote the tightest constraints on the matter dipole amplitude just from the subsample that contains the largest number of sources (lowest flux limit) while still best described by a dipole model.
Of the results listed in Table~1 of \citet{Blake:2002gx}, this is the case at a flux density cut of 25~mJy, where the fit has a reduced $\chi^2$ of unity. Here they find $\mathcal{D}_\mathrm{obs}=(1.1\pm0.3)\times10^{-2}$, a value larger than the kinematic expectation of $4.6\times10^{-3}$ by $2.2\sigma$. This deviation is quite consistent with the significance levels found in later works. 
After all, as pointed out by \citet{Crawford:2008nh}, given the number of NVSS sources above the 15~mJy minimum flux density cut employed by \citep{Blake:2002gx} a 370~km~s$^{-1}$ dipole cannot be detected as non-zero even at $1\sigma$, while \citet{Blake:2002gx} report a $3\sigma$ detection, which is possible only if the measured dipole is considerably larger than the expectation. Hence on closer examination, the first detection of the kinematic matter dipole did \emph{not} in fact confirm the standard expectation and already posed a challenge to the CP.

\subsubsection{The search for independent confirmation}
\label{sec:EBtest:History:IndependentConfirmation}

The following years saw a noticeable increase in attempts to measure the kinematic matter dipole using radio sources. Understandably, the anomalously large dipole amplitude found with NVSS was under much scrutiny and a number of techniques and data sets were employed to uncover possible errors in these first analyses. To this effect, cross-checks were performed by analysing other radio catalogs, individually, or in combination, also to increase the total number of sources and the observed sky fraction. For instance, \citet{Rubart:2013tx} performed measurements with a combination of NVSS and the 325~MHz Westerbork~Northern~Sky~Survey \citep[WENSS;][]{1997A&AS..124..259R}, while \citet{Bengaly:2017slg} and \citet{Singal:2019pqq} considered the 150~MHz TIFR~GMRT~Sky~Survey \citep[TGSS;][]{Intema:2016jhx} in addition to NVSS, and \citet{Colin:2017juj} combined NVSS with the 843~MHz Sydney~University~Molonglo~Sky~Survey \citep[SUMSS;][]{Mauch:2003zh}. In parallel, sample contamination by local sources was studied, e.g.~via cross-matching with low-redshift catalogs~\citep[e.g.][]{Gibelyou:2012ri,Rubart:2014lia,Rameez:2017euv,Bengaly:2018ykb}, and the impact of survey systematics on dipole measurements were modelled such as declination-dependent flux calibration errors~\citep[e.g.][]{Bengaly:2017slg}. Overall, these studies confirmed the indication for the excess dipole amplitude initially found with NVSS alone, yet noticeable scatter among the various results raised suspicions about the presence of unknown systematics in the respective catalogs. Most notably, the dipole measured using TGSS  initially appeared a factor of 5 larger than that of the NVSS~\citep{Bengaly:2017slg}, as discussed below. Some scatter among the values was in fact to be expected, since there was as yet no consensus on the dipole estimator to be used, which led occasionally to biased results in both its direction and amplitude, especially for low source counts, as is the case for all aforementioned radio catalogs (see \S~\ref{sec:UncertaintiesAndSystematics} for a discussion). Ultimately, an independent measurement was needed of the kinematic matter dipole using a diﬀerent dataset.

The \citeauthor{1984MNRAS.206..377E} test is applicable more generally than just with radio sources (\emph{cf.}~\S~\ref{sec:EBtest:Conditions}). A truly independent confirmation therefore would require a measurement using a catalog of extragalactic sources detected in a distinct manner, in a distinct range of frequencies. While such attempts had been undertaken already \citep{Hirata:2009ar,Itoh:2009vc} a catalog of sufficient size and quality to guarantee a significant detection was not yet available. This changed with the advent of the Wide-field~Infrared~Survey~Explorer (WISE) satellite, which allowed for the first uniformly selected samples of quasars to be produced covering most of the sky \citep{2015ApJS..221...12S,Assef:2017apb}. With the NEOWISE and NEOWISE-R continuation surveys \citep{Mainzer:2011ak,2014ApJ...792...30M} the depth and uniformity of WISE data in the quasar-sensitive $W1$ and $W2$ bands increased substantially, resulting in the unWISE \citep{2019ApJS..240...30S} and CatWISE2020 \citep{2021ApJS..253....8M} catalogs. The latter was used by \citet{Secrest:2020has} to construct a sample of nearly 1.4~million quasars covering most of the sky outside of the Galactic plane. They found a dipole amplitude over twice as large as the kinematic expectation, though similar in direction to the CMB dipole, and showed using simulated skies that this could be expected by chance only about one in 2~million times, i.e. a $4.9\sigma$ result --- the most significant at the time. An independent analysis using this sample  confirmed and strengthened this result \citep{Dam:2022wwh}.

The fundamental strength of the WISE quasar-based results is that they were obtained in a way that is systematically independent from radio dipole studies. The WISE data are space- rather than ground-based, at a wavelength 5 orders of magnitude separated from the radio, measured using a completely different instrument design, and employing a scanning strategy with a different principle axis (ecliptic instead of equatorial) and astrophysical foregrounds. Additionally, as discussed in \S~\ref{sec:EBtest:History:StateOfTheArt}, WISE-selected quasars and radio-selected galaxies are nearly entirely separate classes of objects, allowing for a joint statistical analysis. Indeed, the first such analysis by \citet{Secrest:2022uvx} found consistency between the radio galaxy and quasar dipoles and a combined significance of $5.1\sigma$ for the cosmic dipole anomaly.

\subsubsection{The search for an explanation of the anomaly}
\label{sec:EBtest:History:SearchForExplanation}

However there were also results seemingly conflicting with the overall finding of excess matter dipoles in the radio and mid-IR, which we first review.

\begin{figure}
\includegraphics[width=\columnwidth]{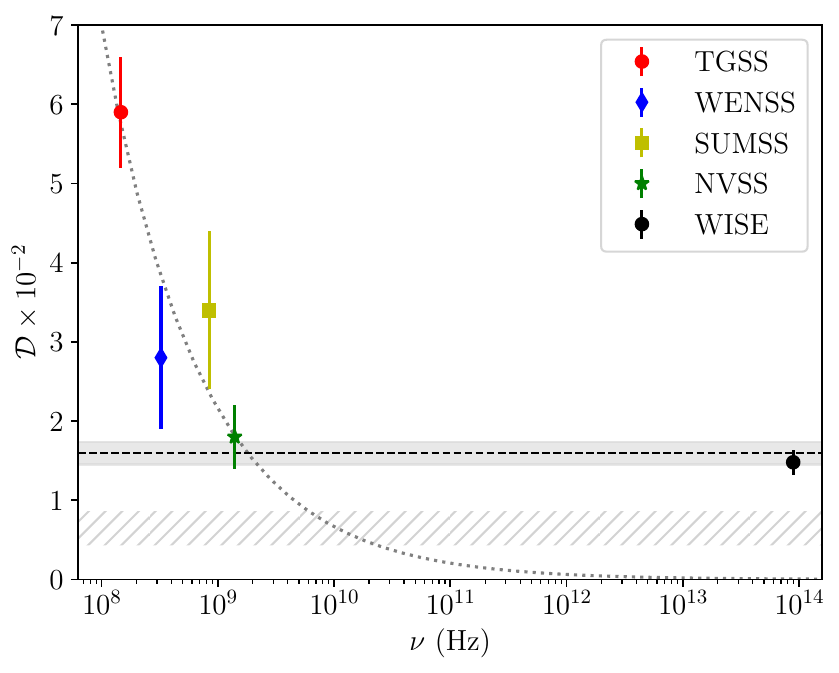}
\caption{TGSS, WENSS, SUMSS, and NVSS dipole amplitudes from \citet{Siewert:2020krp}, compared with the WISE dipole amplitude from \citet{Secrest:2020has,Secrest:2022uvx}. The dotted line indicates the functional frequency dependence proposed by \citet{Siewert:2020krp} using only the radio data, which is however not consistent with the WISE infrared value. The dashed line and shaded interval show weighted arithmetic mean and standard error, excluding the TGSS data point, demonstrating that the data are otherwise generally consistent with \emph{no} frequency dependence of the dipole. The grey hatching indicates the standard kinematic expectation for $1 < x < 2$ and $0.5 < \alpha < 1.5$.}
\label{multifrequency}
\end{figure}

Analysing the TGSS, WENSS, SUMSS, and NVSS \citet{Siewert:2020krp} confirmed the anomalously large dipoles seen in radio studies. However, they  suggested that the dipole has a frequency dependence, which, if true, would imply an astrophysical rather than  cosmological origin. They noted that this is indicated predominantly by the lowest frequency survey used (TGSS), however the functional form proposed by \citet{Siewert:2020krp} is inconsistent with the WISE quasar dipole, which has an amplitude consistent with the NVSS, and the (less well constrained) SUMSS and WENSS dipoles, but not the TGSS --- see Figure~\ref{multifrequency}. The weighted arithmetic mean of the WENSS, SUMSS, NVSS, and WISE dipole amplitudes, using the values from \citet{Secrest:2020has,Secrest:2022uvx}, is $\mathcal{D}_\mathrm{obs} = (1.60\pm0.15)\times10^{-2}$, with a corresponding reduced $\chi^2$ of 1.94 ($p=0.12$ given three degrees of freedom), indicating that the data are consistent with a \emph{constant} dipole amplitude across 5 orders of magnitude in frequency. 

As shown in \citet{Secrest:2022uvx}, the particularly large dipole amplitude of the TGSS can almost entirely be attributed to the flux calibration errors that it is known to have \citep{2017arXiv170306635H}, which when corrected for bring the apparent dipole amplitude from $\mathcal{D}_\mathrm{obs}=6.5\times10^{-2}$ down to $\mathcal{D}_\mathrm{obs}=2.2\times10^{-2}$. \citet{Secrest:2022uvx} also found that further accounting for the dependence between rms noise and source density in the TGSS reduces the dipole amplitude to $\mathcal{D}_\mathrm{obs}=1.4\times10^{-2}$. \citet{Singal:2023wni} however raised a number of objections to the procedure used by \citet{Secrest:2022uvx} and asserted that the TGSS does indeed have a particularly large dipole amplitude. We address this in \S~\ref{sec:UncertaintiesAndSystematics:Biases:FluxCalibrationErrors}, but emphasize that the only datum that currently suggests a frequency dependence of the cosmic matter dipole is the TGSS, which is known to have significant large-scale systematics.

In recent years other catalogs have become available, such as the 887.5~MHz Rapid~ASKAP~Continuum~Survey \citep[RACS;][]{2021PASA...38...58H}, the 3~GHz Karl G. Jansky Very Large Array Sky Survey \cite[VLASS;][]{Lacy:2019rfe}, and a new Gaia-unWISE quasar sample~\citep[Quaia;][]{Storey-Fisher:2023gca}. \citet{Darling:2022jxt} first combined RACS and VLASS to claim a dipole measurement consistent with the standard expectation, although the large uncertainty on his estimate made it equally consistent with the anomalously large dipole amplitude found previously. More importantly, RACS and VLASS individually returned dipoles that grossly disagreed in direction, thus undermining their combined analysis (see \citet{Secrest:2022uvx} for additional discussion). The Quaia sample too promised a cross-check of the quasar dipole, and \citet{Mittal:2023xub} attempted such a measurement, finding a dipole aligned with that of the CMB and consistent with the standard expectation. However large uncertainties mainly due to low source counts in the final subsample prevented a firm conclusion, which was further weakened after they recomputed the source spectral indices \citep{Mittal:2023xub}. Eventually, their model comparison could not distinguish decisively between the preference for or against the standard expectation.

It is natural to extend studies of the matter dipole to higher multipoles, especially in light of the aforementioned anomalies claimed in the CMB at low-$\ell$~\citep{Schwarz:2015cma}. Indeed, this was a prerequisite for estimating the trustworthiness of NVSS even prior to initial matter dipole measurement~\citep{Blake:2001bg,Blake:2002kx,Blake:2004dg}. Within uncertainties this returned smaller-scale correlations that agreed well with the expectation of a $\Lambda$CDM matter power spectrum. In contrast, this was not the case for similar investigations of the TGSS \citep{Tiwari:2019wmj} where issues with flux calibration were found to be able to produce the anomalous power found on larger scales. Along these lines \citet{Tiwari:2022hnf} also investigated the power spectrum of the CatWISE2020 AGN sample of \citet{Secrest:2020has}. After removal of the ecliptic latitude trend found in \citet{Secrest:2020has} this too returned a power spectrum in good agreement with the expectation from standard $\Lambda$CDM at small scales (multipole $\ell \gtrsim 10$), except for, however, the anomalously high dipole.

Despite the ecliptic being the principal observational axis of WISE, a recent study by \citet{Abghari:2024eja} takes issue with the ecliptic latitude correction, simultaneously claiming that the origin of this gradient is unexplained \emph{and} that it is evidence of stellar contamination. These authors  perform simulations generating mock quasar maps with randomly oriented quadrupoles and octupoles allowed to vary in strength up to that of the dipole, showing that the octupole in particular will sufficiently broaden the distribution of dipole amplitudes estimated from these mocks so as to alleviate the bulk of the tension with the dipole estimated using the sample of \citet{Secrest:2020has}. However, even without correcting for the known ecliptic latitude bias, the measured power spectrum does not allow for the octupole to be larger than about half of the dipole, as Figure~9 in \citet{Abghari:2024eja} shows. Simulations constraining the higher multipoles to have powers consistent with the measurements within the uncertainties would have been informative, but this is not what \citet{Abghari:2024eja} did. The nature of the ecliptic latitude bias is described in \citet{Secrest:2022uvx} as arising due to source deblending in deeper imaging at high ecliptic latitudes and somewhat bluer AGNs preferentially scattering into the $W1-W2$ color cut in shallower imaging at low ecliptic latitudes. As the ecliptic axis is set by our Solar System, it is unlikely that the trend is related to stellar contamination as claimed by \citet{Abghari:2024eja}, who also claim that the large Galactic plane mask employed by \citet{Secrest:2020has} was to mitigate stellar contamination. This is untrue as it was in fact to avoid a drop in source density near the Galactic plane (not an increase as would be expected from source contamination). It has long been known that WISE-selected quasar counts decrease closer to the Galactic plane and this effect is due to source confusion \citep[e.g.,][]{2015ApJS..221...12S, Assef:2017apb}. While we certainly encourage further development and characterization of the quasar samples presented in \citet{Secrest:2020has,Secrest:2022uvx}, we do not believe any case has been made for stellar contamination or unexplained higher-order multipole systematics which can significantly affect the quasar dipole anomaly \cite{vonHausegger:2025sub}.

\subsubsection{The state of the art today and summary}
\label{sec:EBtest:History:StateOfTheArt}

The significance of the cosmic dipole anomaly has reached the $5\sigma$ level, using mid-infrared data alone \citep{Secrest:2020has, Dam:2022wwh}, radio data alone \citep{Wagenveld:2023key,Wagenveld:2025ewl}, and in a joint analysis with both mid-infrared and radio data \citep{Secrest:2022uvx}.
Thus it has the same level of statistical significance as the much discussed `Hubble tension' \citep{Riess:2021jrx}. Nevertheless, the cosmic dipole anomaly has not received commensurate attention \citep{Peebles:2022akh}.

While initial observational studies were undertaken almost exclusively with the radio source catalog of the NVSS, the data landscape has changed substantially since, and as a result, research in this field has been rapidly developing. 
For instance, additional radio catalogs, especially those that cover southern declinations such as SUMSS, TGSS, and RACS have offered new opportunity for study. Just as importantly, systematically independent samples of quasars were created using mid-infrared data from WISE \citep{Wright:2010qw}, especially the deep CatWISE2020 catalog \citep{2021ApJS..253....8M} whose analysis propelled this field further. 

An important aspect of having large catalogs of both radio galaxies and quasars is that these two populations are largely independent, corresponding to different evolutionary stages in the histories of galaxies and different accretion AGN accretion modes. Quasars are, by definition, bolometrically dominant AGNs, with intrinsic luminosities exceeding the entire galaxy in which they reside. These luminosities are only possible through the liberation of energy from infalling matter in an accretion disk around a SMBH, with matter-energy conversion efficiencies up to $\sim40\%$ \citep{Blandford:1977ds}. The direct continuum from quasars is therefore thermal, originating in the accretion disk, and largely uncollimated, although the presence of an optically thick, dusty torus creates a viewing angle dependence of the apparent luminosity. The peak of quasar activity is at $z\sim2$, with the number of quasars per comoving volume declining rapidly at low redshift \citep[e.g.,][]{Hopkins:2006fq}. Radio galaxies above a few mJy, on the other hand, are generally powered by non-thermal emission of jets formed in the immediate vicinity of the SMBH and bolometrically sub-dominant AGNs are often radio-luminous. The comoving density of radio sources shows strong redshift evolution, declining rapidly to moderate redshift and exhibiting a radio power-dependent peak significantly lower than that of quasars for all but the most luminous sources \citep[e.g.,][]{Rigby:2011hn}. 

\begin{figure*}[tbh]
\includegraphics[width=\textwidth]{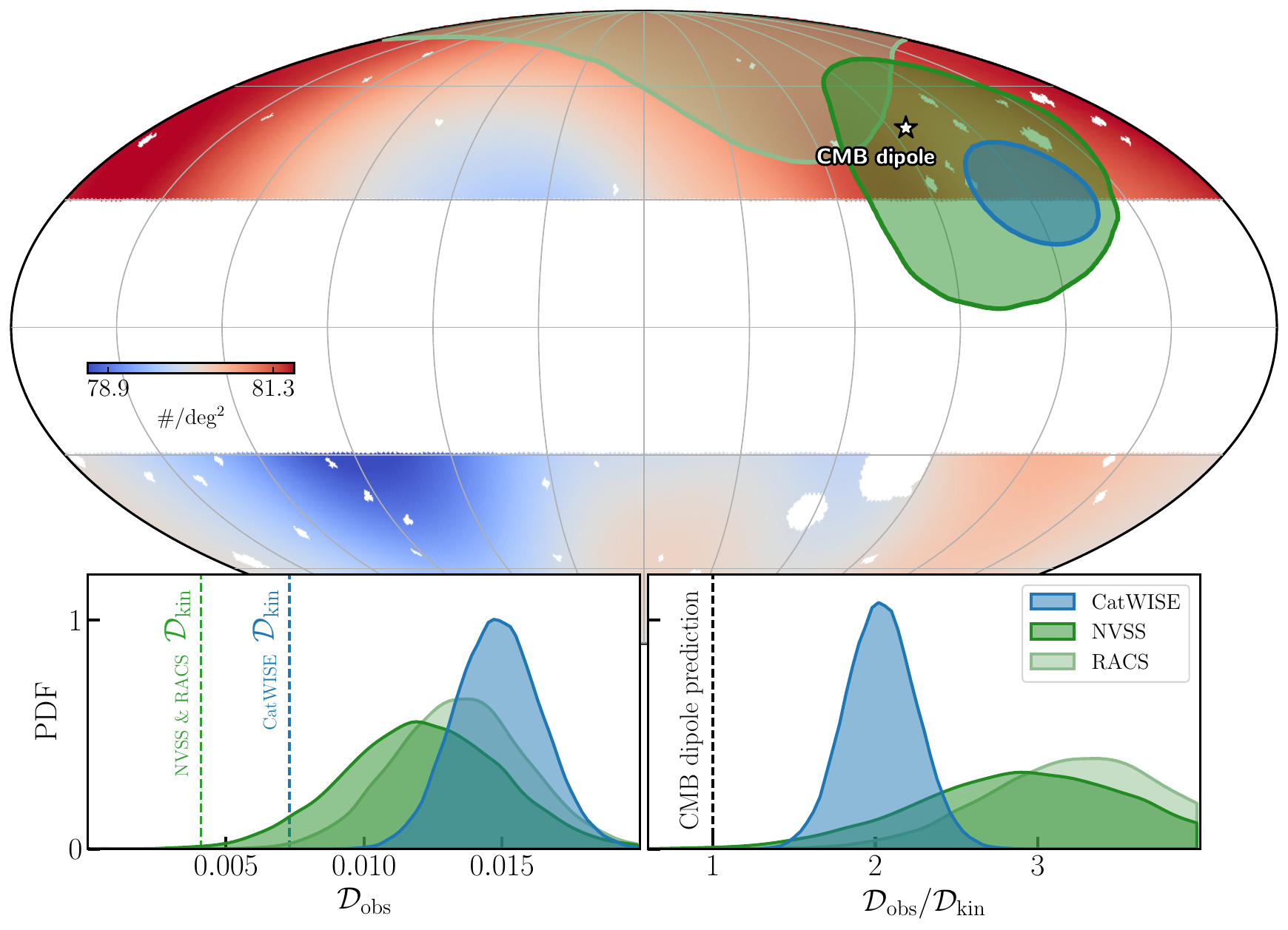}
\caption{Overview of the most significant matter dipole measurements to-date using the CatWISE AGN, NVSS, and RACS-low samples \cite{Secrest:2022uvx,Wagenveld:2025ewl}. Top panel: Contours enclosing 90\% of the posterior probability of the inferred dipole directions plotted onto a smoothed version of the CatWISE AGN number counts. The star marks the CMB dipole direction. Bottom left panel:  Posterior probability densities of the inferred dipole amplitudes $\mathcal{D}_\mathrm{obs}$ compared against their standard expectations, viz. the expected kinematic matter dipole amplitudes $\mathcal{D}_{\rm kin}$ computed for the respective samples. Bottom right panel:  Same as left panel but scaled by the respective kinematic expectations.}
\label{fig:Overview}
\end{figure*}

These differences are accommodated by a model in which radio AGNs and quasars are in different evolutionary stages and accretion modes. Radio AGNs are typically found in massive, red, evolved galaxies while quasars are typically hosted in bluer, less massive, and less clustered galaxies \citep[e.g.,][]{Hickox:2009ak}. In the context of the dipole anomaly, the different populations that make up radio galaxies and quasars have two important implications. Firstly, considering the observational systematic independence of ground-based radio surveys and the space-based WISE infrared survey, the population independence of radio galaxies and infrared-selected quasars makes the anomalous dipole seen in both populations hard to ignore. Indeed, this was the key result of \citet{Secrest:2022uvx}, who performed the first joint dipole analysis of radio galaxies and quasars. The more recent analysis of \citet{Wagenveld:2023kvi} leverages 0.8~million radio galaxies selected using NVSS and RACS data, finding a 4.8\,$\sigma$ disagreement with the kinematic expectation. Using the sample size-weighted $Z$-scores method, 
to combine the \citet{Wagenveld:2023kvi} and \citet{Secrest:2022uvx} results indicates that the dipole anomaly has now reached $\sim6.4$\,$\sigma$ significance. We display these measurements, the most significant made to-date, in Figure~\ref{fig:Overview}, with the CatWISE AGN, NVSS, and RACS-low samples prepared as described in~\citet{Secrest:2022uvx}  and \citet{Wagenveld:2025ewl}, and the inferences  performed with Poisson likelihoods. Note that RACS-low is a newer catalog that deserves further scrutiny for possible systematics, as the CatWISE and NVSS catalogs have been subjected to.

Looking forward, the second result, still nascent, is characterization of the anomaly as a function of cosmological observables such as redshift, clustering, or galaxy bias. Given that the consistency between the radio galaxy and quasar dipoles improves by assuming the CMB dipole to be purely kinematic (rather than having an intrinsic component), as the analysis by \citet{Secrest:2022uvx} suggested and the work of \citet{Wagenveld:2025ewl} further supported, the matter dipole may be due to an over-density of radio galaxies and quasars in the CMB frame, on scales way beyond that which can be accommodated in the standard cosmology \citep[][]{Peebles:2022akh}. Redshift tomography~\citep{vonHausegger:2024fcu} of millions of objects on near-full sky scales would illuminate the moderate-redshift universe sufficiently to make progress on this issue.

\section{Uncertainties and Systematics}
\label{sec:UncertaintiesAndSystematics}

The observed matter dipole may be described as a vector $\textbf{d}_\mathrm{obs}$ resulting from the vector sum of the kinematic dipole $\textbf{d}_\mathrm{kin}$, the clustering dipole $\textbf{d}_\mathrm{cls}$ due to an imbalance of matter in our local Galactic neighbourhood, and a ``shot noise'' dipole $\textbf{d}_\mathrm{SN}$ arising from a random distribution of finite source counts. There may also be a contribution to the observed dipole due to survey systematics that affect the calibration of the data. In this review, we consider all contributions to the  dipole other than the kinematic one to be systematics. By carefully accounting for or mitigating these, one can compare the observed dipole to the theoretical expectation, allowing for a consistency test of the null hypothesis (i.e., that the Universe is exactly described by FLRW). We discuss these systematics below.

\subsection{The clustering dipole}
\label{sec:UncertaintiesAndSystematics:ClusteringDipole}

Extragalactic surveys that sample only the immediate comoving volume ($r\lesssim150h^{-1}$~Mpc) due to a shallow flux limit or are biased towards more cosmologically evolved objects will be dominated by local clustering, which can be described by a density field parametrized by spherical harmonics with power in all multipoles, including the dipole ($\ell=1$). For a typical observer in the concordance $\Lambda$CDM cosmology, this ``clustering dipole'' $\mathcal{D}_\mathrm{cls}$ should follow the power spectrum of mass density perturbations, and is computable given knowledge of the object redshift distribution ${\rm d}N/{\rm d}z$ and the linear bias $b$ of the objects with respect to the total underlying mass density \citep[e.g.,][]{Gibelyou:2012ri,Rameez:2017euv}. Under the null hypothesis, accurate knowledge of the object redshift distribution and bias yields a rigorous estimate of $\mathcal{D}_\mathrm{cls}$, allowing for determination of the importance of the clustering dipole when determining the consistency of the observed dipole with the kinematic prediction. 

The clustering dipole $\mathcal{D}_\mathrm{cls}$ in a sample of sources as seen by a typical observer in the concordance $\Lambda$CDM cosmology is related to the power spectrum $P(k)$ of (dark) matter density perturbations via:
\begin{align}
    \mathcal{D}_\mathrm{cls} = \sqrt{\frac{9}{4\pi}C_1} ,
\end{align}
\noindent where
\begin{align}
    C_\ell = \frac{2}{\pi} \int_0^{\infty} f_\ell(k)^2 P(k) k^2 \mathrm{d}k .
\end{align}
\noindent Here the filter function $f_\ell(k)$ includes the redshift-dependent linear bias $b(z)$  of the sources with respect to the dark matter:
\begin{align}
    f_\ell(k) = \int_0^{\infty} b(r(z)) j_\ell(kr)f(r)dr ,
\end{align}
\noindent where $j_\ell$ is the spherical Bessel function of order $\ell$ and $f(r)$ is the probability distribution for the comoving distance $r$ to a random object in the survey:
\begin{align}
    f(r) = \frac{H(z)}{H_0 r_0}\frac{\mathrm{d}N}{\mathrm{d}z} ,
\end{align}
\noindent normalised such that $\int_0^{\infty} f(r) \mathrm{d}r = 1$ and $\mathrm{d}N/\mathrm{d}z$ is the redshift distribution of the observed objects.

The theoretical estimate of the expected clustering dipole amplitude for a random observer is useful to quantify the null hypothesis for a given catalog. Dipole measurements specifically in shallow, low-redshift catalogs can further inform this expectation~\citep[e.g.][]{Yoon:2014daa,Alonso:2014xca}. Likewise low-redshift catalogs can assist with removing low-redshift-contamination~\citep[e.g.][]{Colin:2017juj,Rameez:2017euv,Schwarz:2018swk,Bengaly:2018ykb,Cheng:2023eba,Oayda:2024hnu}. We discuss below some astrophysical considerations relevant for assessing the clustering dipole.

\subsubsection{Quasars}
\label{sec:UncertaintiesAndSystematics:ClusteringDipole:Quasars}

The redshift distribution of quasars, which are typically detected in optical spectroscopic surveys, is known with high confidence and peaks at $z\sim1-2$, with almost no quasars found at $z < 0.1$ where clustering starts becoming significant. It is important to distinguish here between quasars and active galactic nuclei (AGNs) more generally. The latter are found in much higher abundance in local systems because they are often detected using methods that do not rely on bolometric dominance over the host galaxy, such as emission line ratio diagnostics or detection at hard X-ray energies ($>$~few keV), and objects may be classified in the literature as AGNs based on any evidence for nuclear activity in the host galaxy, at any bolometric luminosity. For example, the recent and final release of the Milliquas catalog \citep{2023OJAp....6E..49F} contains 860,100 objects identified as quasars and 47,044 AGNs. While 17\% of the AGNs have $z< 0.1$, almost none of the quasars are below this redshift. 

While the majority of spectroscopic quasars are known from the SDSS/BOSS Quasar Catalog (DR16Q), which covers 28\% of the celestial sphere, producing a sample of quasars covering closer to $\sim50\%$ of the sky is feasible using mid-infrared color criteria. This technique leverages the fact that quasars produce strong power-law emission between $\sim1$~$\mu$m and $\sim10$~$\mu$m where inactive galaxies either show a strong decline in emission along the Rayleigh-Jeans tail of older stellar populations, or a notable decrement of emission between the Rayleigh-Jeans tail and an increase in emission due to star formation-heated dust at longer wavelengths. The source of mid-infrared emission in quasars is generally thought to be dust heated to the sublimation limit ($\sim1500$~K) by the central engine, much hotter than the characteristic dust temperature of star-forming regions, producing distinctive mid-infrared color, especially notable in the $[3.4]-[4.6]$ (W1$-$W2) color using data from the Wide-field Infrared Survey Explorer \citep[WISE;][]{Wright:2010qw}. Critically, the mid-infrared color of quasars also cleanly separates quasars from nearly all Galactic stars, circumventing the requirement for costly and currently unavailable all-sky spectroscopy \citep{2015ApJS..221...12S}.

Defining a cosmologically useful sample of quasars covering enough of the sky to test the dipole is therefore possible in principle, and indeed was first carried out by \citet{Secrest:2020has}. By matching to objects with both SDSS and BOSS spectroscopic coverage along Stripe~82, which was included in the Dark~Energy~Survey footprint, \citet{Secrest:2020has} obtained spectroscopic redshifts for 61\% of WISE-selected quasars, with the remaining 39\% having optical-to-infrared color consistent with heavily reddened quasars, which have a similar redshift distribution as unobscured quasars \citep[see, e.g.,][]{Petter:2023imf}. With the redshift distribution of WISE-selected quasars confidently determined, \citet{Secrest:2020has} estimated that the clustering dipole of these objects is negligibly small. This was confirmed by \citet{Tiwari:2022hnf} who observed that the angular power spectrum of CatWISE is well-fitted by the $\Lambda$CDM expectation on small scales (multipole $\ell \gtrsim 10$) and determined directly by extrapolation to $\ell=1$ that $\mathcal{D}_\textrm{cls}=0.81 \times 10^{-3}$. A recent study \cite{vonHausegger:2025sub} shows that the quadrupole, octupole etc.~too are all much smaller than the dipole and consistent with $\Lambda$CDM, so the anomalously high dipole cannot be explained away as due to the presence of a high octupole as claimed by \citet{Abghari:2024eja}. 

\subsubsection{Radio galaxies}
\label{sec:UncertaintiesAndSystematics:ClusteringDipole:RadioGalaxies}

Unlike quasars, which are bolometrically dominant AGNs luminous at infrared wavelengths, most radio galaxies are known through the detection of non-thermal synchrotron radiation produced via shocks, in either star-forming galaxies in the nearby universe or (much more commonly above a few mJy) radio-bright AGNs at moderate redshift. Consequently, radio galaxies are often extremely faint at optical wavelengths where large spectroscopic surveys are carried out. For example, the famous radio galaxy Messier~87, at a Virgo cluster distance of 16.5~Mpc, would have an apparent SDSS $r$-band magnitude of $\sim21$ at $z=0.5$, well below the spectroscopic targeting flux limit of SDSS. The situation is further complicated by the fact that many radio sources detected at $\sim1$~GHz (e.g., in the NVSS, SUMSS, or RACS surveys) are AGN-driven radio lobes at considerable angular offsets from their source galaxy, making counterpart identification especially difficult. Nonetheless, the CENSORS survey \citep{Rigby:2011hn} of NVSS objects contains redshift measurements for 76 objects above the 15~mJy flux density cut typically employed in dipole studies, of which \emph{none} have $z<0.1$ where contamination by the clustering dipole starts to become significant in the concordance $\Lambda$CDM model. However, the binomial confidence interval still allows for $\sim5\%$ of NVSS objects to have $z<0.1$ (CL~$=95\%$), and moreover this survey was taken from a small, $\sim6$~deg$^{2}$, patch of the sky so cosmic variance may be significant. Nonetheless, indirect estimates \citep[e.g.,][]{Ho:2008bz} suggest that there is no underestimated population of radio sources at low redshift \citep[although see][for a dissenting view]{Cheng:2023eba}. 

In several investigations of the radio galaxy dipole, an attempt was made to remove local clustering sources, starting with \citet{Blake:2002gx}, who removed NVSS sources associated with the IRAS PSCz Catalog and the Third Reference Catalog of Bright Galaxies (RC3). By binning in redshift, they showed that the contribution of the clustering dipole to the radio galaxy dipole measurement drops to minimal levels by $z = 0.03$. \citet{Tiwari:2013vff}, using a similar spherical harmonic-based estimator, also tested the NVSS dipole with the PSCz and RC3 sources removed, finding $\mathcal{D}_\mathrm{obs}=1.1\times10^{-2}$ and $\mathcal{D}_\mathrm{obs}=1.5\times10^{-2}$ for flux density cuts of 10~mJy and 20~mJy, respectively, about 60\% larger than the values reported in \citet{Blake:2002gx} for the same cuts. \citet{Tiwari:2013vff} further explored removing sources from the 2MASS Large~Galaxy~Atlas, the 2MASS~Redshift~Survey (2MRS), and the 2MASS~Photometric~Redshift~Catalog (2MPZ), which were not available at the time of \citet{Blake:2002gx}. These catalogs, especially the 2MPZ that is based largely on the deeper 2MASS~Extended~Source~Catalog (XSC), increase completeness in the local universe and to higher redshifts, further removing possible clustering dipole contamination by local sources. \citet{Tiwari:2013vff} found that the NVSS dipole is essentially unaffected by removal of these additional sources, yielding $\mathcal{D}_\mathrm{obs}=9.6\times10^{-3}$ and $\mathcal{D}_\mathrm{obs}=1.3\times10^{-2}$ for 10~mJy and 20~mJy flux density thresholds. Similarly, \citet{Colin:2017juj} determined the radio galaxy dipole using a nearly full sky sample of sources from the NVSS and SUMSS. They cross-correlated their sample of radio galaxies with 2MRS and additionally explored the effect of removing the supergalactic plane, which may contribute significant power from local clustering. For all cuts applied, however, the amplitude and direction of the radio galaxy dipole remained stable, with the amplitude significantly larger than the 370~km~s$^{-1}$ kinematic expectation. 

Given these results, while it should be acknowledged that the exact redshift distribution of radio sources (from catalogs such as NVSS) is not as well known as that of quasars, it has been convincingly shown that removal of the most important contributors to the local clustering dipole has little effect on the radio galaxy dipole. 

\subsection{Biases}
\label{sec:UncertaintiesAndSystematics:Biases}

\subsubsection{Flux calibration errors}
\label{sec:UncertaintiesAndSystematics:Biases:FluxCalibrationErrors}

All catalogs created from real data have position-dependent sensitivity differences due to  systematics inherent in the survey pattern or instrument setup. For example in the case of NVSS, the switch from the Very Large Array D to DnC configuration at high and low declinations resulted in loss of sensitivity at the nominal $\sim4$~mJy flux limit of the survey. In the case of WISE, source counts at the faint end are limited by confusion noise, which causes a loss of sensitivity for objects above some flux threshold in regions with higher source density, such as near the Galactic plane or regions of deeper coverage closer to the ecliptic poles. These sensitivity differences are effectively statistical noise and can be mitigated by simply raising the flux cut, typically to 10~mJy or higher for radio catalogs \citep[e.g., Fig.~1 in][]{Wagenveld:2023kvi}, or parameterizing the effect as a weighting function as was done for WISE \citep{Secrest:2020has,Secrest:2022uvx} as well as NVSS and RACS \citep{Wagenveld:2023kvi}. Additional systematics may be due to 
loss of completeness near bright sources, which additionally raise the \emph{effective} flux uniformity limit of the catalog. The first step, therefore, in constructing a catalog suitable for the Ellis-Baldwin test is to bin source counts as a function of the principal directions (declination, ecliptic latitude, Galactic latitude, and---if the clustering dipole is a concern---supergalactic latitude), and then test for dependencies choosing the flux cut and mask employed accordingly.

\begin{figure}
\includegraphics[width=\columnwidth]{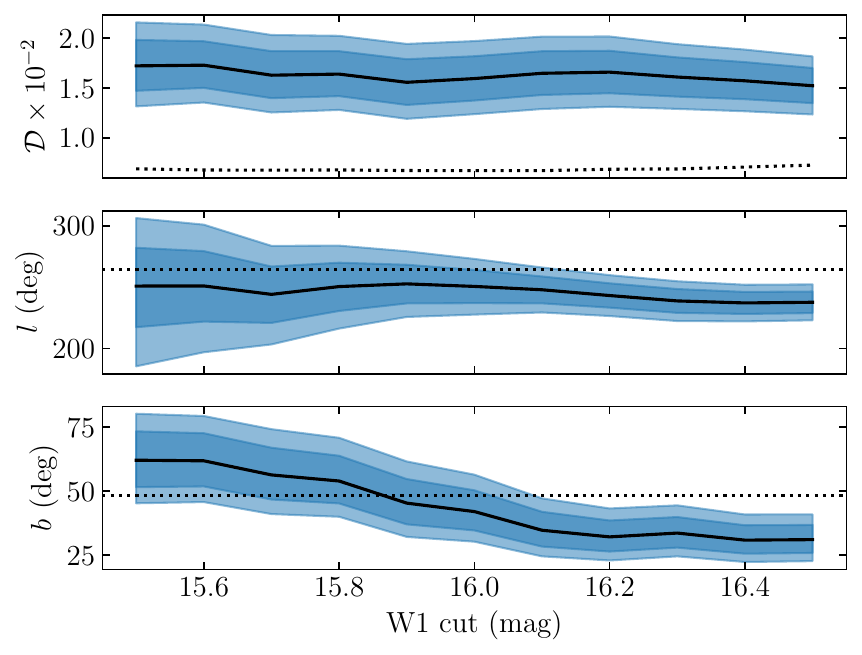}
\caption{The variation with flux cut of the dipole amplitude and its direction for WISE-selected quasars; the dotted lines show the kinematic expectation corresponding to the CMB dipole. It is seen that the  amplitude remains stable (and about twice as high as the CMB-based expectation) while the direction remains close to that of the CMB dipole. The light and dark blue regions indicate the 5--95 and 16--84 percentiles of the posteriors after marginalizing over all other parameters, while the solid black line shows the median.}
\label{wise_dipole_vs_w1cut}
\end{figure}

A catalog of moderate redshift objects for which the local clustering dipole is negligible should exhibit a dipole that is stable in both direction and amplitude with increasing flux cut, if it is indeed purely kinematic. In Figure~\ref{wise_dipole_vs_w1cut} we show such a test on the CatWISE2020 sample (marginalizing over the ecliptic latitude trend) which demonstrates the stability of its dipole amplitude and direction. However, if there is an \emph{intrinsic} component to the dipole and the flux cut has some correlation with redshift, then the measured dipole may drift somewhat due to real astrophysical effects as the flux cut is increased. Hence while it disfavors unaccounted for observational systematics, this test should not be a strict prerequisite for determining if the true dipole has been detected, given that the reason for the disagreement of the amplitude with the kinematic expectation remains unknown.

Although statistical error can be mitigated by raising the flux cut or masking problematic regions, the same is not true for systematic (position-dependent) flux calibration errors, which manifest as source counts variance that persists across a wide range of flux cuts. Flux calibration errors can be either corrected in the analysis or appropriately masked. It is likely that the VLA~Sky~Survey \citep{Lacy:2019rfe}, as used e.g. by \citet{Darling:2022jxt}, suffers from calibration errors, at least for decl.~$<-15^\circ$. \citet{Darling:2022jxt} did note this issue and masked below decl.~$<-15^\circ$, but it is likely that this is insufficient to make VLASS suitable for the Ellis-Baldwin test, at least for the first epoch, as argued by \citet{Secrest:2022uvx}.

The TGSS catalog notably exhibits a dipole several times larger than any other radio catalog employed for the Ellis-Baldwin test \citep{Bengaly:2017slg} and is the critical data point that justifies the claim by \citet{Siewert:2020krp} that the dipole has a frequency dependence. \citet{Secrest:2022uvx}, on the other hand, ascribe the large TGSS amplitude to flux calibration errors, an effect they demonstrate by recalibrating the TGSS using NVSS. \citet{Singal:2023wni} argues that, contrary to the findings of \citet{Secrest:2022uvx}, the startlingly high dipole in the TGSS catalog cannot be ascribed to flux calibration errors, because the TGSS dipole points in a direction similar to the CMB dipole, and that correcting for flux calibration errors as \citet{Secrest:2022uvx} did artificially makes the number density of sources in TGSS match that in NVSS, resulting in a similar dipole amplitude. 

In fact, the TGSS is well known to have position-dependent flux calibration errors ranging between $\sim0.6-1.6$ \citep{2017arXiv170306635H}, so a measurement of its dipole should \emph{prima facie} not be taken as an accurate estimate of the radio source dipole, especially given the significant excess power between $2 < \ell < 30$ observed in this catalog \citep[][see also their Section 3.2.1]{2019A&A...623A.148D}. Indeed, \citet{Siewert:2020krp}, who used the apparently larger dipole in the TGSS to argue for a frequency dependence of the dipole amplitude, found that the TGSS dipole points $39^\circ$ away from the CMB direction using a 100 mJy cut as in \citet{Singal:2023wni}, the same offset found using the code employed by \citet{Secrest:2022uvx},\footnote{\url{https://zenodo.org/record/6784602}} although the directions differ by $11^\circ$. Correcting for flux calibration errors using the NVSS moves the dipole direction only by $16^\circ$ despite reducing its amplitude from $\mathcal{D}_\mathrm{obs}=6.5\times10^{-2}$ to $\mathcal{D}_\mathrm{obs}=2.2\times10^{-2}$. 

Second, flux calibration at very low frequencies ($\lesssim500$~MHz) is notoriously difficult, due to the presence of radio frequency interference as well as the bright Galactic synchrotron foreground. Flux calibration of a 150~MHz survey such as the TGSS is inherently much more difficult than that of a 1.4~GHz survey such as the NVSS, and indeed despite being over 25 years old there have been, to our knowledge, no papers pointing out significant flux calibration issues with the NVSS. It is therefore not unreasonable to use the NVSS to correct the fluxes of the TGSS, and because the NVSS footprint overlaps almost completely with that of the TGSS it may be the only option if the TGSS is to be used for dipole work. Allowing for the use of the NVSS as a flux reference sample, correcting TGSS can be done by requiring that the average spectral across all sources be the typical value for optically thin synchrotron: $\alpha=0.75$. Correction is done on a sky pixel by pixel basis, choosing sky pixels large enough to contain enough sources that their mean spectral index is likely to be close to the fiducial value, but small enough to allow determination of a position-dependent flux correction factor. Employing this procedure, \citet{Secrest:2022uvx} showed unambiguously that the uncorrected TGSS source density is correlated with the observed per-pixel mean $\alpha$ and that using the observed per-pixel mean $\alpha$ effectively corrects for artefacts in the TGSS source density map (their Figure 3). Given these considerations, we conclude that the balance of the evidence disfavours the much larger dipole apparent in the TGSS. 

As illustrated above, flux calibration errors can pose a serious challenge to an unbiased measurement of large-scale clustering.  In anticipation of future data, such effects could be studied by simulating source catalogs with corresponding flux uncertainties, as was done, e.g., by \citet{Bengaly:2017slg}.  Naturally such issues deserve individual scrutiny for each galaxy sample.

\subsubsection{Dipole estimators}
\label{sec:UncertaintiesAndSystematics:Biases:Estimators}

In addition to biases in the calibration of source properties, signal estimation itself can be biased for a given catalog. Various estimators have been used in the literature to measure dipole amplitude and direction, each of which comes with strengths and weaknesses. With increased attention to this issue many biases have now been accounted for.We provide a short overview of the estimators employed, their potential biases, and modern developments.\\

Initially, the so-called linear estimator~\citep{Crawford:2008nh}, a simple vector sum over the source positions, was commonly employed due to its simplicity~\citep[e.g.][]{Singal:2011dy,Rubart:2013tx}. However, when applied to a masked sky it required often non-trivial corrections depending on the particular shape of the survey mask. It was believed that samples whose sky coverage was not point symmetric with respect to the zero point had to be masked additionally so as to symmetrize their footprint, in order to avoid a bias in amplitude~\citep{Rubart:2013tx}. Use of the linear estimator on the NVSS, for example, would require that the $-40^\circ$ declination limit be mirrored by cutting all sources above $+40^\circ$, $\sim20\%$ of the catalog. However, even so, directional biases could not be prevented, requiring further corrections which only for simple masks had analytical and therefore computationally efficient form~\citep{2015PhDT.......262R,Siewert:2020krp}. For these reasons, despite its computational efficiency, the linear estimator has since been replaced with other methods.

Alternatively, the decomposition of the sky into spherical harmonics also allowed the estimation of the dipole, however again only if no mask was applied. To account for the breaking of the orthogonality of the harmonics on a masked sky, and the concomitant bias, higher multipoles had to be considered and evaluated simultaneously, involving sky simulations to estimate the harmonics' correlation~\citep[e.g.][]{Blake:2002gx}. This too appeared to prove impractical at the time although today modern methods can help alleviate some of the computational load to handle masked data~\citep[e.g.][]{Wolz:2024dro}.

In light of these biases, template fits were invoked instead~\citep{Hirata:2009ar}, so-called quadratic estimators, simple $\chi^2$~\citep[e.g.][]{Rubart:2013tx} or least-squares statistics~\citep[e.g.][]{Secrest:2020has} over a pixelized sky, minimised for the dipole direction and amplitude. Correlation among multipoles in the presence of a mask is not generally circumvented with these methods either, but the accompanying biases are negligible if the dominant multipoles on the sky are all included in the fit. For instance, if the data contained more structure than a simple dipole, a quadrupole for example, due either to systematics or source clustering, then a pure dipole model can lead to a bias. To account for this, some authors have extended their template fits to also include quadrupole terms, or even templates of particular contaminants, to constrain the presence of such terms. If the model is well-specified, however, these methods are unbiased in the limit of a large number of sources.

Initial evaluations of the significance of any departure from the null hypothesis were almost exclusively Frequentist, which required for millions of ``null skies'' to be generated and fit, from which the $p$-value may be calculated. However, early minimisation methods were rather inefficient, limiting their feasibility for the evaluation of $p$-values~\citep[e.g.][]{Bengaly:2018ykb}. For this purpose, least-squares estimators are especially suited, due to their analytical solution allowing for millions of mock skies to be fit~\citep[e.g.][]{Secrest:2020has}.

Nevertheless, biases may remain: if the source counts per unit solid angle are non-Gaussian for the sky model such quadratic models may be misspecified. This is strictly true even for an ideal catalog of uncorrelated sources, which will have Poisson-distributed counts. Therefore there will be significant departures from the estimators' Gaussian assumption if source counts are low or if they are correlated, leading to an overestimation of the dipole amplitude. Typically, this becomes noticeable for source counts $\lesssim600,000$, i.e.~for source densities of $\lesssim 15\,{\rm deg}^2$, so this affected estimates of the NVSS dipoles slightly, but not those of the CatWISE AGN.

Nowadays, many of above issues have been settled: The evaluation of Poisson likelihoods including appropriate priors on the template parameters avoid biases in the recovered parameters, even for low source counts~\citep[e.g.][]{Wagenveld:2023kvi,Dam:2022wwh}; Frequentist methods have been replaced with Bayesian inference and model comparison~\citep[e.g.][]{Mittal:2023xub}, avoiding the need to generate large numbers of mock skies for the evaluation of significances, trivially including templates of higher multipoles~\citep[e.g.][]{Oayda:2024voo}; joint evaluation of combined catalogs at different frequencies does not require density matching if their likelihoods can be treated independently~\citep[e.g.][]{Wagenveld:2023kvi,Oayda:2024hnu}; even complicated masks do not prevent the measurement of an underlying source density modulation~\citep[e.g.][]{Wagenveld:2023key,Wagenveld:2024qhn}; and while the responsibility remains to bias-test likelihoods and inference pipelines with mocks, efficient MCMC samplers~\citep[e.g.][]{Hoffman:2011ukg} allow for computationally tractable analyses.

Lastly, correlation in source counts may occur for a variety of reasons, including double-lobed sources in radio catalogs and, more broadly, dual AGN sources in catalogs~\citep[e.g.][]{Blake:2004dg}. Source counts may also become correlated if the presence of other sources affects the probability of a given source being successfully extracted from image data (e.g., the effects of source deconvolution). Consequently, the within-cell source counts may more accurately be described by the negative binomial (or ``compound Poisson'') distribution, as in e.g., \citet{Siewert:2019poc}. Using this distribution, every model likelihood takes on an additional parameter $p$ that varies as $0<p<1$, with the source counts distribution converging to Poissonian in the limit of $p\rightarrow1$. In practice, it is unlikely that marginalizing over this additional parameter will increase the uncertainties on the other model parameters enough to meaningfully affect interpretation, but we recommend including it to account for the effect of correlated source counts.\footnote{We thank Lukas~B\"{o}hme and Dominik Schwarz for this comment.}

\subsubsection{Weighted estimators}
\label{sec:UncertaintiesAndSystematics:Weights}

The effect described by \citeauthor{1984MNRAS.206..377E} explains the modulation of number counts of sources across the sky by a dipole, due to special relativistic aberration and Doppler boosting and depending on the source properties. It is possible to extend these considerations by predicting a dipole modulation of weighted number counts, perhaps with the goal to disentangle the contribution of underlying causes. For instance, \citet{Nadolny:2021hti} consider number counts weighted by combinations of flux, size, and redshift, to distinguish an intrinsic from a kinematic contribution to the number count dipole, as discussed in \S~\ref{sec:LCDM}. But even just as a means of seeking a consistency with the unweighted \citeauthor{1984MNRAS.206..377E} test, several authors~\cite{Singal:2011dy, Rubart:2013tx, Darling:2022jxt} had already considered measuring dipoles in flux-weighted number counts, with methods sometimes referred to as a flux weighted estimators. While such variations offer  opportunities, additional considerations are warranted, which we exemplify by discussing flux weighting.

First and foremost, when weighting by the flux, the brightest few sources will dominate the measured signal. For example, using the NVSS sample from \citet{Secrest:2022uvx}, the 9\% of sources with flux densities above $S_{1.4~\mathrm{GHz}} > 110$~mJy contribute more to a simple weighted vector addition than the remaining 91\% of fainter sources. For the WISE-selected quasars, the corresponding numbers are $\sim310,000$ sources out of 1.6~million (19\%). Second, because of the enormous range spanned by the luminosity distances of astronomical objects, objects at cosmological distances suitable for the \citet{1984MNRAS.206..377E} test will naturally be fainter, and Galactic contaminants, sources contributing to local clustering, or even catalog artifacts may be more prevalent at brighter flux levels. For example, by making a map of the 2MASS Extended Source Catalog (XSC) sources, which form a flux-complete sample of galaxies out to $z\sim0.1$, we find a correlation coefficient of 0.056 with the $S > 10$~mJy NVSS sample used in \citet{Secrest:2022uvx}, indicating that 5.6\% of NVSS sources are below $z < 0.1$. However, restricting to the 9\% of sources with $S > 110$~mJy that dominated the total flux of the NVSS, the fraction goes up to 14\%. Moreover, the theoretical expectation (corresponding to the CMB dipole) for the dipole amplitude of flux-weighted number counts can be quantified only by making further assumptions. For example, \citet{Singal:2011dy} requires that the same power-law behavior of the number counts holds for all objects in the survey (rarely the case for real data, see, e.g., Figure~1 of \citet{Singal:2019pqq}), whereas \citet{1984MNRAS.206..377E} requires only that the power-law behaviour is applicable very close to the  flux limit of the survey.

Cautious considerations of this kind are also called for when constructing weights involving galaxy size or redshift. Nevertheless, the large amount of upcoming data from large-scale surveys in the near future should provide opportunity to consider studying the cosmic dipole anomaly in such different ways.

\subsection{Summary of theoretical issues}
\label{sec:UncertaintiesAndSystematics:Theory}

The \citet{1984MNRAS.206..377E} test requires data which fulfills specific conditions (\S~\ref{sec:EBtest:Conditions} \& \S~\ref{sec:EBtest:Sensitivity}) and potential deviations from these requirements must be addressed either prior to undertaking the test or by adjusting the method with which the dipole is  measured.

The  kinematic matter dipole amplitude (\ref{eq:Dkin}) differs from the kinematic CMB dipole amplitude (\ref{eq:cmbdipole}) in its prefactor $\left[2+x(1+\alpha)\right]$, which contains information about the astrophysical properties of the sample. In \S~\ref{sec:EBtest:SR} $x$ was defined as the power-law index of a fit to the (redshift-integrated) integral source counts (\ref{eq:x}), \emph{at the flux threshold} $S_*$. While \citeauthor{1984MNRAS.206..377E} assumed a perfect power law to hold for any value of $S_*$, this is \textit{not} a prerequisite for undertaking the test successfully; the power-law approximation need only hold at the applied flux limit of the catalog (\citet{vonHausegger:2024jan} and \S~\ref{sec:EBtest:Conditions}). Modeling deviations from a perfect power law over a larger range of $S_*$ than is actually required according to Eq.(\ref{eq:deltaSoverS}), as proposed by \citet{Tiwari:2013vff} and used by \citet{Siewert:2020krp},  can in fact mispredict $\mathcal{D}_{\rm kin}$.

The possible influence of astrophysics on the prefactor in Eq.(\ref{eq:Dkin}) sparked a recent discussion about its accuracy when $\alpha$ and $x$ vary with redshift along the line of sight. This was said to lead to changes in Eq.(\ref{eq:Dkin}) which were termed ``theoretical systematics''~\citep{Dalang:2021ruy,Guandalin:2022tyl}; to calculate these apparently requires full knowledge of the sources' luminosity function. These arguments were however erroneous in that they did not incorporate the crucial requirement (\S~\ref{sec:EBtest:Conditions}) that the spectral index $\alpha$ describes only those sources that are very close to the flux limit \citep{vonHausegger:2024jan}. There are thus no ``theoretical systematics'' in the \citeauthor{1984MNRAS.206..377E} formula.

Nevertheless, the redshift-dependent formulation (see \S~\ref{sec:EBtest:GR}) by \citet{Maartens:2017qoa} provides a good starting point to study potential corrections to Eq.(\ref{eq:Dkin}) when additional cuts must be made on quantities affected by special relativistic boosts, e.g. while performing redshift-tomography or to accommodate sample selection beyond just a flux cut~\citep{vonHausegger:2024fcu}. Indeed, future data  (\S~\ref{sec:Future}) will offer more opportunities to understand the kinematic dipole signal by also including fluxes, redshifts, sizes etc.~in extensions of \citeauthor{1984MNRAS.206..377E}'s original idea. We have briefly discussed caveats regarding such approaches  (\S~\ref{sec:UncertaintiesAndSystematics:Weights}), however in principle they can disentangle intrinsic contributions e.g.~clustering (see \S~\ref{sec:UncertaintiesAndSystematics:ClusteringDipole}) from kinematic contributions to the measured matter dipole~\citep[cf.][]{Nadolny:2021hti}.

Not all data sets, however, satisfy the requirements of \citet{1984MNRAS.206..377E}; in particular spectral lines may prevent the straightforward application of Eq.(\ref{eq:Dkin}) to observations in the optical, where the power law approximation of the spectrum can break down if there are not enough objects in the range of redshifts to adequately smooth out the observed SED. Moreover unforeseen selection effects may emerge in new data that may ultimately require numerical implementations of the test, viz.~forward-modeling the raw observations in a dedicated framework.\\

\section{Future Galaxy Surveys}
\label{sec:Future}

Looking ahead, we await results from several forthcoming cosmological data sets which will allow the redshift dependence of the cosmic matter dipole anomaly to be determined. We provide below a brief discussion of relevant observables and forecasts.

Of the galaxy catalogs that have so far been used to measure the cosmic matter dipole, radio observations, due to their robust detections of AGN at  $z\sim1$, make up the largest fraction. Besides the  NVSS, samples from the TGSS, WENSS, and SUMSS have also been studied, both individually and in combination. In preparation for the Square Kilometre Array (SKA), radio catalogs by pathfinder missions have been employed as well, e.g. VLASS and RACS. Most recently, data from the MeerKAT~\citep{2016mks..confE...1J} absorption line survey \citep[MALS;][]{2016mks..confE..14G} was prepared for a dipole measurement~\citep{Wagenveld:2023key}.

While the CatWISE2020~\citep{2021ApJS..253....8M} quasar catalog \citep{Secrest:2020has} is so far the only sample based on mid-IR data that has been used for matter dipole measurements, combinations of various samples can distil novel high-redshift catalogs via synergistic methods. For instance, the Gaia-unWISE quasar sample~\citep{2019MNRAS.489.4741S} combines the unWISE catalog with Gaia measurements for AGN classification. Similarly, using the updated Gaia DR3 release it was attempted to use the Quaia catalog~\citep{Storey-Fisher:2023gca} for a dipole measurement~\citep{Mittal:2023xub}.

Many of these studies contributed  to the development of new tools and understanding regarding optimal dipole study design, from which future studies will benefit. However, all the above named samples, except for the CatWISE2020 AGN sample, have insufficient source counts to detect the expected kinematic dipole above the $\sim1\sigma$ level, as emphasized by \citet{2025RSPTA.38340027S}. Moreover due to the increased uncertainties, while many works report agreement of their findings with the expected kinematic dipole, they fail to note that there is equal or even better agreement with the anomalously large dipole, as unveiled at the highest significance by \citet{Secrest:2020has}. Moreover  the presence of systematic issues in some of the catalogs used (e.g.~TGSS, VLASS) prevent reliable measurements on large scales altogether. We must await large-volume surveys in the near future, that can address the cosmic dipole anomaly conclusively.

Fortunately, several upcoming surveys will almost certainly reveal important new information about the cosmic dipole anomaly, in large part due to the increase in source counts that will allow sensitivity at the level of the kinematic expectation (currently only possible with WISE data, as noted above) and in part due to sampling not just radio galaxies and quasars but normal galaxies as well, yielding a third, mostly independent population of moderate-redshift matter that, importantly, is also less biased relative to the dark matter distribution. These surveys will however have their own systematic limitations. With the enormous increase in source counts the dominant source of error will be systematic, such as error from uncertainties in Galactic reddening---a significant problem for near-infrared data and especially data at visual wavelengths. Nonetheless, these surveys are likely to deepen our understanding of the nature of the cosmic dipole anomaly. We discuss below the most relevant near future surveys of cosmologically distant sources, viz. SPHEREx, Euclid, Rubin-LSST, and SKA.

\subsection{SPHEREx}
\label{sec:Future:SPHEREx}

SPHEREx \citep[Spectro-Photometer for the History of the Universe, Epoch of Reionization, and Ices Explorer,][]{SPHEREx:2014bgr} which was launched this year will carry out the first all-sky, visual to near-IR spectral survey. SPHEREx will survey the sky with 102 photometric bands spanning 0.75--5\,$\mu$m and achieve a $5\sigma$ depth of $\sim19$~AB between 0.75--5\,$\mu$m and $\sim18$~AB at 5\,$\mu$m,\footnote{\url{https://spherex.caltech.edu/page/survey}} corresponding to WISE magnitude limits of $W1\lesssim16$ and $W2\lesssim15$ in the Vega system. With $6^{\prime\prime}$ spectral pixels, deeper data would be limited by confusion noise, so it is unlikely that SPHEREx will achieve a significant depth improvement over current WISE data in terms of quasar counts. However, the 102 channels of SPHEREx correspond to a spectral resolution of $R\sim40$, allowing for redshift tomography~\citep{vonHausegger:2024fcu} of both quasars and galaxies. SPHEREx is expected to provide accurate redshifts for $\sim100$~million galaxies out to $z < 1$ \citep{SPHEREx:2018xfm}, so it should be possible to accurately deconvolve the effect of local clustering from the kinematic dipole, with transition to moderate redshift.

\subsection{Euclid}
\label{sec:Future:Euclid}

The Euclid satellite~\cite{Euclid:2024yrr} will have surveyed almost 15,000~deg$^2$ of the sky at near-IR wavelengths by the time of its third data release. Up to 25 million galaxy redshifts between  $0.9 < z < 1.8$ will be measured spectroscopically during its baseline mission, in addition to roughly a billion-and-a-half photometric redshifts, across a distribution with median redshift $z\sim1$~\cite{Euclid:2019clj}. These data will allow for redshift tomography \citep{vonHausegger:2024fcu} and be complementary to SPHEREx; Euclid will probe higher redshifts and have 20 times better angular resolution, mitigating source confusion and allowing for measurements of galaxy morphologies. Euclid will eventually shed light on the redshift range where the dipole anomaly arises, its structure (e.g., if it exhibits $\ell>1$ components), and whether it is a function of galaxy bias (dependent on source type). 

The availability of Euclid Quick Data Release \citep[Q1;][]{Euclid:2025rvk} data allows a broad forecast of the power of Euclid to understand the dipole anomaly. Using objects selected from the Euclid deep fields, we select likely galaxies with $H_\mathrm{E} < 24$, the nominal completeness limit by Year~6 of the Euclid~Wide~Survey \citep[EWS;][]{Euclid:2021icp}, by removing sources with point-like photometry. 
Using the $J_\mathrm{E} - H_\mathrm{E}$ values for power-law SEDs provided by \citet{Euclid:2022vkk}, we find $\alpha \sim -3.6\,(J_\mathrm{E} - H_\mathrm{E})$ ($S_\nu \propto \nu^{-\alpha}$). For galaxies close to the completeness limit, $\langle \alpha \rangle \sim -1.0$, and we also find $dN(>S)/dS \propto S^{-0.8}$, predicting a kinematic dipole amplitude of $\mathcal{D}_\mathrm{kin} \sim 2.5\times10^{-3}$. The predicted dipole amplitude due to shot noise is $\mathcal{D}_\mathrm{SN} \sim 3 \sqrt{f_\mathrm{sky}/N}$, where $f_\mathrm{sky}$ is the fraction of the sky observed and $N$ is the number of sources observed. Given the 14,514~deg$^{-2}$ footprint by year 6 of the EWS and $\sim131$,000 galaxies per square degree down to $H_\mathrm{E} < 24$~mag \citep{Euclid:2024few}, there should be 1.9 billion galaxies observed across 35\% of the sky, which implies $\mathcal{D}_\mathrm{SN} \sim 4\times10^{-5}$, i.e. 60 times smaller than the kinematic prediction. The Euclid dipole error is therefore likely to be dominated by systematic effects, such as the clustering dipole and the effects of Galactic reddening.

Given the depth of the EWS, the dipole due to local clustering should be negligible. Using the photometric redshifts from the Euclid Quick Data Release \citep{Euclid:2025vsf}, we estimate that only $\sim0.4\%$ of galaxies detected by Euclid have $z<0.1$, where the clustering dipole starts becoming significant compared to the kinematic prediction. It is likely, therefore, that uncertainties in Galactic reddening will dominate the error. While the EWS footprint largely avoids the Galactic plane, careful assessment of the effect of Galactic reddening uncertainties should nonetheless be made, as the typical practice of assuming a uniform total-to-selective extinction ratio (e.g., $R_V = 3.1$) may not be safe, given the sensitivity of dipole work.

\subsection{Rubin/LSST}
\label{sec:Future:LSST}
The Legacy Survey of Space \& Time~\citep[LSST;][]{LSST:2008ijt} by the Vera C.\ Rubin Observatory is a synoptic astronomical survey with a $\sim20,000$~deg$^2$ footprint. A systematic scan of the celestial sphere will be performed over ten years, leading to the largest astronomical catalog ever compiled with $\sim4$ billion galaxies out to $z\lesssim3$ \citep{LSSTScience:2009jmu}.
As with Euclid, the high depth and wide sky coverage of LSST means that studies of the dipole anomaly will be limited by systematic error, and the systematic error that will likely present the greatest challenge is calibrating for Galactic reddening, viz. the variation in the total-to-selective extinction ratio $R_V$, known to systematically vary over large areas as $\sigma_{R_V} = 0.18$ with apparently no correlation with color excess $E(B-V)$ \citep{2016ApJ...821...78S}. Given a reddening coefficient $A_\lambda / A_V = 0.395$ in the $y$ band \citep{2019ApJ...877..116W}, this means that dipole studies with LSST will likely be limited to $E(B-V)$ about ten times smaller than when using WISE data for the same level of systematic error contribution. The WISE-based samples used in previous studies have $E(B-V) < 0.3$ for nearly all objects, given the Galactic plane mask used, so this suggests that, without calibration of position-dependent $R_V$, LSST-based studies will need to use windows with $E(B-V) < 0.03$ to mitigate systematic error, only 18\% of the sky, or about 3600~deg$^2$ given LSST's footprint. The usual practice of assuming a static $R_V=3.1$ may thus significantly limit LSST's usefulness in dipole studies, although the deep photometric redshifts from LSST will be valuable when working with data less impacted by Galactic reddening such as Euclid. Fortunately, the extraordinary depth and photometric precision of the final, co-added data will allow measurement of $R_V$ to high Galactic latitude \citep[][Section~7.5.2]{LSSTScience:2009jmu}, potentially mitigating much of the systematic error due to reddening, so an accurate measure of the dipole may indeed be possible. Our forecast for LSST is that, as with Euclid, a thorough treatment of the uncertainties due to Galactic extinction must be made if the wide sky footprint of LSST is to be effectively used to study the dipole of moderate redshift galaxies.

\subsection{Square Kilometre Array}
\label{sec:Future:SKA}

The Square Kilometre Array Red Book~\cite{SKA:2018ckk} highlights the potential of the Phase 1 of SKA data to shed further light on the cosmic dipole. Covering over an order of magnitude in frequency, from $151\,{\rm MHz}$ in the LOW configuration to $1.4\,{\rm GHz}$ in the MID configurations, the SKA is anticipated to deliver the highest-sensitivity large-volume radio measurements to date over $3\pi$ of the sky. For instance, flux limited samples of around 100 million radio sources~\citep{Schwarz:2015pqa} are expected to become available with the SKA1 baseline design (over 1 billion with SKA2) collected at the MID frequency of NVSS, enabling the dipole to be characterised at unprecedented statistical precision, while the availability of HI redshifts should allow all sources at $z \leq 0.5$ to be removed~\citep{Bengaly:2018ykb} thus suppressing the clustering dipole.

SKA's ability to detect faint sources, down to flux levels of $\mathcal{O}(\mu{\rm Jy})$ also affects the composition of its catalogs. In contrast to NVSS, which operated at flux limits of $\mathcal{O}(10)$\,mJy, the SKA will detect a large number of star-forming galaxies (SFGs), which tend to lie at lower redshifts than the $z\sim 1$ AGN otherwise studied. The influence thereof on dipole measurements is already becoming apparent in pathfinder studies such as the recently compiled catalog of MALS sources~\cite{Wagenveld:2023key,Wagenveld:2024qhn}. It will be interesting to compare results of the measured radio dipoles also for higher flux thresholds, effectively increasing the median redshift of the sources. With threshold-dependent integral source count scaling,\footnote{See Table 3 in \citet{SKA:2018ckk}, where the index $x$ in Eq.(\ref{eq:x}) is denoted as $\alpha_{\rm mag}$.} comparison to different dipole expectations $\mathcal{D}_{\rm kin}(S_*)$ will offer further opportunities to understand the nature of the matter dipole.

\section{Concluding Remarks}
\label{sec:conclusion}

In recent years, the Cosmological Principle which has been an article of faith for a century and on which the standard $\Lambda$CDM model rests, has come under increasing scrutiny \cite[see,][]{Aluri:2022hzs}. This is mainly because the data necessary to check it observationally has finally become available and also because cracks have begun appearing meanwhile in the `concordance' $\Lambda$CDM model, e.g. the `Hubble tension' \cite[see,][]{Abdalla:2022yfr,Perivolaropoulos:2021jda}.

The cosmic dipole anomaly which strikes at the heart of the CP is an even more serious problem as it applies to all FLRW models. Whereas there are solutions to Einstein's equations which do not assume any symmetries viz. the Szekeres models \citep[e.g.][]{Celerier:2024dvs}, these have a much larger number of parameters and the community has been reluctant to consider these when the simple FLRW-based $\Lambda$CDM model has been so successful in fitting data. However one of its `simple' parameters is the Cosmological Constant $\Lambda$ which, interpreted as the energy density of the quantum vacuum, would require fine-tuning of two unrelated terms to at least 60 decimal places to enable the Universe to exist in its present form. It is clear that simplicity is in the eye of the beholder.

Nevertheless there is a natural tendency, based on
the premise that extraordinary claims require extraordinary evidence, to retain prior beliefs until the weight of the contrary evidence forces a change. Moreover a transition to a new paradigm requires a new theoretical framework and it is fair to say that no compelling explanation has yet been presented for the cosmic dipole anomaly.

In a prescient essay, \citet{1988ASPC....4..344G} had noted the then emerging discordance between peculiar velocities and the mass distribution inferred from galaxy catalogs and speculated whether the anomalous bulk flow could in fact be due to a superhorizon perturbation in the curvature. There can be a mismatch between the reference frames in which radiation and matter are isotropic in such a `tilted universe' and the idea was developed further by \citet{Turner:1991dn}. However a detailed examination by \citet{Domenech:2022mvt} showed that while such a perturbation can reduce the CMB dipole amplitude, it has little effect on the matter dipole. Hence the good alignment of the two must be accidental if our real velocity corresponds to the latter and the smaller amplitude of the CMB dipole is due to such a cancellation by an isocurvature mode. This therefore is not a satisfactory explanation of the dipole anomaly (within the FLRW paradigm).

While more complex general relativistic models might describe a Universe with a dipole anisotropy, they lack clear physical motivation. Severe bounds on anisotropic expansion are also set by  CMB temperature and polarisation data \cite{Saadeh:2016sak}. Going beyond FLRW, the spatially closed Kantowski-Sachs Universe and the open axisymmetric Bianchi type III Universe have been considered \cite{Constantin:2022dtj}, and more general Thurston geometries \cite{Awwad:2022uoz}, as well as the Szekeres models \cite{Celerier:2024dvs}, in order to compute observationally relevant quantities. Another suggestion has been of a axially isotropic, tilted Bianchi V/VII$_\text{h}$ cosmology which accommodates a cosmic flow \cite{Krishnan:2022qbv}. An interesting suggestion is that large-scale anisotropy may emerge from the growth of non-linear structure and be described by locally rotationally symmetric Bianchi cosmologies that allow for both anisotropic expansion and large-scale bulk flow \citep{Anton:2023icm,Anton:2024swg}. To establish which, if any, of these ideas or others yet to be proposed, is on the right track, will require further observations and serious engagement with the `fitting problem' in cosmology \cite{Ellis:1987zz}. 

\acknowledgments
We are grateful to many colleagues for correspondence and discussions on this topic, in particular David Alonso, Reza Ansari, Chris Blake, Camille Bonvin, Charles Dalang, Lawrence Dam, Harry Desmond, Guillem Domenech, Ruth Durrer, George Ellis, Pedro Ferreira, Geraint Lewis, Roy Maartens, Jim Peebles, Dominik Schwarz, Prabhakar Tiwari, Wilbur Venus and Jonah Wagenveld. Jacques Colin inspired us to work on this topic and participated in our first analyses. We thank the anonymous referees for their critical reading and suggestions which improved this review. 

\bibliographystyle{apsrmp4-1}
\bibliography{refs}

\end{document}